\documentclass[11pt]{article}
\usepackage[T1]{fontenc}
\usepackage[utf8]{inputenc}
\usepackage{newpxtext}  
\usepackage{bm,amsmath,amssymb,mathtools,graphicx}
\usepackage{textcomp,gensymb,units,upquote}
\usepackage[a4paper]{geometry}
\usepackage{epsfig}
\usepackage[parfill]{parskip}
\usepackage[matrix,arrow,color]{xy}
\usepackage[usenames]{color}
\usepackage{mathrsfs}
\usepackage[font={small}]{caption}
\usepackage{subcaption}
\usepackage[export]{adjustbox}
\usepackage{float}
\usepackage{wrapfig}
\usepackage{microtype}
\usepackage{slashed}
\usepackage{braket}
\usepackage[pdftex]{hyperref}
\usepackage{cite}

\input{xy}
\xyoption{all}

\input{epsf}

\makeatletter

\usepackage{setspace}

\usepackage{lscape}

\catcode`@=11
\renewcommand{\section}
{\@startsection{section}{1}{0pt}{\medskipamount}{\medskipamount}{\large\bf}}
\makeatletter\renewcommand{\subsection}
{\@startsection{subsection}{2}{\z@}{-3.25ex plus -1ex minus -.2ex}
{1.5ex plus .2ex}{\it }}

\numberwithin{equation}{section}
\catcode`@=12

\newcommand{\ban}{\begin{eqnarray}}
\newcommand{\ean}{\end{eqnarray}}

\newcommand{\Tr}{{\rm Tr}}
\newcommand{\tr}{{\rm tr}}

\newcommand{\cW}{{\cal W}}

\newcommand{\cM}{{\cal M}}

\newcommand{\cB}{{\cal B}}

\newcommand{\cH}{{\cal H}}
\newcommand{\cA}{{\cal A}}

\newcommand{\cO}{{\cal O}}

\newcommand{\cR}{{\cal R}}

\newcommand{\cF}{{\cal F}}
\newcommand{\cD}{{\cal D}}

\newcommand{\cK}{{\cal K}}

\newcommand{\cV}{{\cal V}}
\newcommand{\cU}{{\cal U}}

\newcommand{\sfa}{{\mathsf{a}}}
\newcommand{\sfb}{{\mathsf{b}}}

\newcommand{\sfq}{{\mathsf{q}}}
\newcommand{\sfp}{{\mathsf{p}}}

\newcommand{\sfy}{{\mathsf{y}}}
\newcommand{\sfx}{{\mathsf{x}}}

\newcommand{\complex}{{\mathbb C}} 
\newcommand{\zed}{{\mathbb Z}} 
\newcommand{\real}{{\mathbb R}} 

\def\e{{\,\rm e}\,}

\def\ii{{\,{\rm i}\,}}
\def\dd{{\rm d}}

\def\beq{\begin{equation}}
\def\bee{\begin{equation}}
\def\eeq{\end{equation}}
\def\bea{\begin{eqnarray}}
\def\eea{\end{eqnarray}}
\def\bd{\begin{displaymath}}
\def\ed{\end{displaymath}}

\newcommand{\Cint}{\int\kern-10.5pt-\kern7pt}

\newcommand{\be}{\begin{equation}}
\newcommand{\ee}{\end{equation}}
\newcommand{\bal}{\begin{align}}
\newcommand{\eal}{\end{align}}

\newcommand\fverbit{\egroup\item[\fbox{\unhbox\pippobox}]}
\newbox\pippobox

{

\def\be{\begin{equation}}
\def\ee{\end{equation}}
\def\bea{\begin{eqnarray}}
\def\eea{\end{eqnarray}}

\begin{document}

\begin{titlepage}
\setcounter{page}{1}

\vskip 5cm

\begin{center}

\vspace*{3cm}

{\Large GRAVITATIONAL ALGEBRAS AND APPLICATIONS TO \\[8pt] NONEQUILIBRIUM PHYSICS}

\vspace{15mm}

{\large Michele Cirafici}
\\[6mm]
\noindent{\em Dipartimento di Matematica, Informatica e Geoscienze, \\ Universit\`a di Trieste, Via A. Valerio 12/1, I-34127,
 \\ Institute for Geometry and Physics \& INFN, Sezione di Trieste,  Trieste, Italy 
}\\ Email: \ {\tt michelecirafici@gmail.com}

\vspace{15mm}

\begin{abstract}

\noindent This note aims to offer a non-technical and self-contained introduction to gravitational algebras and their applications in the nonequilibrium physics of gravitational systems. We begin by presenting foundational concepts from operator algebra theory and exploring their relevance to perturbative quantum gravity. Additionally, we provide a brief overview of the theory of nonequilibrium dynamical systems in finite dimensions  and discuss its generalization to gravitational algebras. Specifically, we focus on entropy production in black hole backgrounds and fluctuation theorems in de Sitter spacetime.

\end{abstract}

\vspace{15mm}

\today

\end{center}
\end{titlepage}


\tableofcontents


\section{Introduction}

The last years have seen remarkable progress in our understanding of quantum gravity. Key insights arising from the AdS/CFT correspondence and careful investigations of the euclidean gravity path integral have led to a series of novel results, the most striking of which is perhaps the computation of the Page curve for an evaporating black hole \cite{Almheiri:2019psf,Penington:2019npb,Penington:2019kki,Almheiri:2019qdq}. Central to these developments is the Quantum Extremal Surface paradigm  \cite{Engelhardt:2014gca}, which can be seen as statement about how quantum gravity degrees of freedom are organized. These new insights can potentially lead to an understanding of how gravity and its thermodynamic aspects emerge from a microscopic description. 

Equilibrium statistical mechanics is one of the triumphs of twentieth century physics. Via an understanding of microscopic degrees of freedom, the study of macroscopic thermodynamic quantities opens a window on the quantum world. This theme carries on in contemporary research on quantum gravity, where thermodynamic quantities, in particular the entropy, provide a guide to interpreting the theory at a microscopic level. The main difference with the triumphs of the last century is the lack of experimental inputs. In the absence of experimental guidance, internal consistency and mathematical rigour are now key players. 

In ordinary quantum field theory entropies and density matrices are difficult to define. This is due to a universal divergence associated to the infinite entanglement of the vacuum state. By using holography one can associate operator algebras to certain backgrounds \cite{Leutheusser:2021qhd,Leutheusser:2021frk}. When perturbative quantum gravity effects are taken into account the algebra of observables becomes of a peculiar kind, known as a type $\mathrm{II}$ von Neumann factor \cite{Witten:2021unn}. For these algebras entropies and density matrices, as well as other thermodynamic quantities can be constructed rigorously. These algebras appear for example when studying quantum fields in a black hole background \cite{Chandrasekaran:2022eqq} or in de Sitter space \cite{Chandrasekaran:2022cip} and they play a crucial role in discussing thermodynamical properties in the presence of gravity. Similar results hold also for other backgrounds \cite{Aguilar-Gutierrez:2023odp,AliAhmad:2023etg,Bahiru:2023ify,Boruch:2024kvv,Engelhardt:2023xer,Faulkner:2024gst,Gesteau:2023hbq,Gomez:2023wrq,Jensen:2023yxy,Kolchmeyer:2023gwa,Krishnan:2023fnt,Kudler-Flam:2023qfl,Leutheusser:2022bgi,Penington:2023dql,Penington:2024sum,Xu:2024hoc,Gesteau:2024rpt,Kolchmeyer:2024fly}. Furthermore they play a role when discussing quantization of constrained systems or the dynamics of observers in gravitational backgrounds \cite{AliAhmad:2024wja,AliAhmad:2024eun,AliAhmad:2024vdw,AliAhmad:2024saq,Chen:2024rpx,DeVuyst:2024pop,DeVuyst:2024uvd,Fewster:2024pur,Geng:2024dbl,Hoehn:2023ehz,Kudler-Flam:2024psh,Klinger:2023tgi,Klinger:2023auu,Witten:2023qsv,Witten:2023xze}. We shall refer to these algebras as gravitational algebras. 

The purpose of this note is to give a quick overview of some developments concerning gravitational algebras, with particular regards to out-of-equilibrium physics. There are by now excellent reviews on the topic, concentrating on equilibrium aspects \cite{Witten:2018zxz,Witten:2021jzq,Sorce:2023fdx,Sorce:2023gio,Sorce:2024pte}. While there are several mathematical tools to explore the equilibrium physics of operator algebras, nonequilibrium dynamics is significantly less understood. There are however a few results on finite dimensional quantum systems, discussed for example in the reviews by D. Ruelle \cite{RuelleRev} and Jak\v{s}i\`{c} and Pillet \cite{JP1}. In this review we will explain how these results extend to the case of gravitational algebras. For a different perspective concerning non-equilibrium aspects of gravitational algebras see also \cite{Kudler-Flam:2023qfl,Chen:2024rpx,Kudler-Flam:2024psh}

In this note we will focus more on the general ideas than on the technical details, for which we refer the reader to \cite{Cirafici:2024jdw,Cirafici:2024ccs}. The main points we would like to explain are how to induce nonequilibrium dynamics by couping the system to reservoirs, and fluctuation theorems. In the first setup we interpret the gravitational algebra appearing in the eternal black hole in AdS as a quantum dynamical system and discuss abstractly how this system can be perturbed by the coupling to external reservoirs. This coupling can induce typical out-of-equilibrium behavior, such as the presence of nonequilibrium steady states and entropy production. The second example we want to discuss concerns de Sitter spacetime, where we show how to adapt the two-times measurement scheme to study dynamical fluctuations. We discuss general forms of out-of-equilibrium fluctuation theorems and discuss some aspects specific to type $\mathrm{II}$ algebras.

This note is organized as follows. In Section \ref{GandH} we will quickly introduce the main geometries we will focus on. Section \ref{OA} discusses some aspects of the theory of operator algebras, in particular focussing on modular theory and quantum dynamical systems. In Section \ref{GA} we will introduce gravitational algebras as they appear in the background of the eternal black hole and in de Sitter. Section \ref{NE} reviews the nonequilibrium dynamics of quantum dynamical systems in finite dimensions while Section \ref{QT} will discuss nonequilibrium dynamics in the context of gravitational algebras.  The Appendix \ref{spectral} contains a few details about the spectral theorem.

\section{Gravity and holography} \label{GandH}

In this Section we will quickly introduce the two main geometries that we will focus upon, the eternal black hole in AdS and de Sitter spacetime. Both geometries have a similar thermodynamic behaviour, they both have horizons and associated entropies.

\subsection{Black Holes in AdS} \label{BHAdS} 

The AdS/CFT correspondence states that any theory of quantum gravity on a spacetime which asymptotically looks like $AdS_{d+1} \times M$, for some manifold $M$, can be described in terms of a relativistic conformal field theory on $\real \times S^{d-1}$. As part of the correspondence between the two theories the symmetries match on both sides and the two Hilbert space are identified. Furthermore the boundary limit of local bulk fields determine operators in the boundary via the so-called extrapolate dictionary, which refers to the behavior of bulk fields near the boundary of AdS, where their scaling is determined by the conformal dimension of the corresponding boundary operator. There are by now several excellent reviews on the topic, we refer the reader to \cite{Harlow:2018fse} for a review closer to the scope of this note.

According to the rules of the duality, when the rank of the CFT $N$ is large the bulk has a geometrical description in terms of Einstein gravity coupled to matter. An important class of operators in the CFT are the single trace primary operators $\cW_i$. Their $k$-point functions scale like $N^{2-k}$.

We are interested in the situation where a black hole is present in the bulk AdS space. If we impose reflecting boundary conditions at infinity and the black hole is big enough, then Hawking radiation is reflected back into the bulk. As a result the black hole reaches thermal equilibrium with its surroundings and will never evaporate completely. This is the so-called eternal (AdS-Schwarzschild) black hole
\be
\dd s^2 =  - f(r) \dd t^2 + \frac{1}{f (r)} \dd r^2 + r^2 \dd \Omega \, ,
\ee
where $f (r)$ is a certain function which vanishes linearly at the black hole horizon. The full geometry has the form of a (non-traversable) wormhole (the black hole interior) connecting two asymptotic regions, labelled L for 'left' and R for 'right', as well as a past and future singularity. This has to be contrasted with a black hole formed by gravitational collapse which only has a future singularity.

This black hole is dual  to two copies of the CFT entangled in the thermofield double (or Hartle-Hawking) state
\be
\ket{\Psi} = \frac{1}{\sqrt{Z}} \sum_i \e^{- \beta E_i / 2} \ket{E_i}_L \ket{\overline{E}_i}_R \, .
\ee
The thermofield double state is a purification of the original thermal state, in the sense that taking a partial trace of the density matrix $\ket{\Psi}\bra{\Psi}$ over one of the two copies of the doubled system gives the density matrix of a thermal state. See \cite{Maldacena:2001kr} for a detailed discussion.

The Hamiltonian acting on the full system is the difference between the Hamiltonians of the left and right copy of the CFT, $H = H_R - H_L$ which is dual to the bulk Hamiltonian. The latter generates time evolution via the isometry $\partial_t$, where the time coordinate $t$ runs forward on the right boundary and backwards on the left.

The black hole's Bekenstein-Hawking entropy is proportional to the area of the horizon \cite{Bekenstein:1972tm,Hawking:1975vcx}
\be \label{BHentropy}
S = \frac{A}{4 G} \, .
\ee
The asymptotic observer sees the vacuum in the near horizon region like a thermal state at finite temperature. This is because the asymptotic Hamiltonian which generates time translations, looks like a boost near the horizon, basically due to outgoing geodesics diverging exponentially near the horizon (a manifestation of the redshift effect). Since any state in QFT looks like the vacuum at short distances, the state of the quantum fields immediately outside the horizon looks thermal. We refer the reader to \cite{Wall:2018ydq} for a more detailed review of black hole thermodynamics.

Physically it is natural to interpret the Bekenstein-Hawking entropy as the logarithm of the number of states of an Hilbert space. But which Hilbert space? The key idea, sometimes called the central dogma \cite{Almheiri:2020cfm} of black hole physics, is that this is the Hilbert space needed to describe the black hole by an observer who is outside the horizon. In other words an observer which remains outside the black hole sees it as a quantum system with as many as \eqref{BHentropy} degrees of freedom.  

\subsection{De Sitter} \label{dS}

The other geometry we will consider is de Sitter spacetime, the maximally symmetric solution of Einstein equations with a positive cosmological constant. In global coordinates its metric is
\be
\dd s^2 = - \dd \tau^2 + \ell^2 \cosh \left( \frac{\tau}{\ell} \right) \dd \Omega_{d-1}^2 \, ,
\ee
which describe a sphere $S^{d-1}$ which has minimum radius $\ell$ at $\tau = 0$ and expands both towards the future and backwards into the past. The cosmological constant is related to the radius by $\Lambda = d (d-1) /  2 \ell^2$.

An inertial observer sits on a point of $S^{d-1}$, say the north pole, and travels along a geodesic. The \textit{static patch} is the intersection of the region which can causally affect the observer with the region which can be causally affected by the observer. In the static patch the observer is surrounded at all times by a null surface, the cosmological event horizon.

As in the black hole case, to this horizon we can associate a temperature: the observer sees Hawking radiation emanating from the cosmological horizon \cite{Gibbons:1977mu}. This can be shown, for example, by Wick rotating to imaginary time. The condition that the resulting metric is smooth requires the Euclidean time coordinate to be periodic. From its period one can read a temperature of $T = \frac{1}{2 \pi \ell}$, where $\ell$ is the de Sitter radius. Note that this temperature is fixed by the de Sitter geometry. By computing the Euclidean path integral as a sum over all compact smooth geometries but in the leading saddle-point approximation, and by interpreting the Euclidean action as proportional to a free energy, one can identify the Gibbons-Hawking entropy of de Sitter space. The result is $S = \frac{A}{4 G}$ where $A$ is the area of the cosmological horizon.

As in the case of black holes, this entropy has the physical interpretation as measuring the logarithm of a certain Hilbert space. In this case this is presumably the Hilbert space that an observer in the static patch needs to account for all those degrees of freedom which lie behind the cosmological horizon, and are therefore lost to her.

\subsection{Generalized Entropy}

The interpretation of \eqref{BHentropy}, as well as its de Sitter counterpart, as an entropy follows from an analogy with classical thermodynamics. This is however somewhat puzzling, since classically a black hole has very few degrees of freedom, such as its mass or its spin. What is lacking in the discussion is the microscopic/statistical interpretation of the Bekenstein-Hawking result as an entropy.

If there is matter outside the black hole then its entropy should be properly taken into account and the relevant quantity is the generalized entropy
\be \label{GenEntropy}
S_{\mathrm{gen}} = \frac{A}{4 G} +  S_{\mathrm{out}}
\ee
where $S_{\mathrm{out}}$ is the von Neumann entropy of the quantum fields outside of the black hole horizon. The latter contains the quantum excitations which constitute the Hawking radiation. The concept of generalized entropy was introduced by Bekenstein to account for the fact that one can reduce the outside entropy by letting matter fall inside the black hole. The generalized entropy \eqref{GenEntropy} obeys a generalized second law of thermodynamics, as it cannot decrease under time evolution \cite{Bekenstein:1972tm}.  The entropy term $S_{\mathrm{out}}$ is UV divergent due to the infinite entanglement of the quantum vacuum, as discussed for example in \cite{Witten:2018zxz}. Remarkably this divergence is proportional to the black hole area and can be absorbed in the first term making the generalized entropy a UV-finite quantity, at leading order. A finite quantity which is cutoff independent is expected to give us information about quantum gravity.

\section{Operator algebras and modular theory} \label{OA} 

In this Section we review some aspects of operator algebras with a particular view towards modular theory and quantum dynamical systems. Some standard textbooks, close to the spirit of this review, are \cite{BR1,BR2} as well as the reviews \cite{RuelleRev,Sorce:2023fdx,Witten:2018zxz}.

\subsection{Some background material}

Here we collect some useful background results. We will be mostly concerned with operators acting on Hilbert spaces. Recall that a Hilbert space $(\cH , \braket{ \ \vert \ } )$, or $\cH$ for short, is a linear space equipped with an inner product $\braket{ \ \vert \ } \, : \cH \times \cH \longrightarrow \complex$ which is linear in the first argument, obeys $\braket{x \vert y } = \overline{\braket{y \vert x}}$ and $\braket{x \vert x} \ge 0$ for every $x,y\in \cH$. The inner product naturally defines a norm on the Hilbert space given by $\Vert x \Vert = \braket{x \vert x}^{1/2}$.

An orthonormal basis of $\cH$ is a sequence of elements $(v_i)_{i \in I}$ such that $\braket{v_i \vert v_j} = \delta_{ij}$ and such that linear combinations of its elements are dense. For a separable Hilbert space the index set $I$ is countable and its cardinality is the dimension of the Hilbert space.

An operator $\sfa$ is \textit{bounded} if
\be
\Vert \sfa \Vert = \sup_{x \neq 0} \frac{\Vert \sfa x \Vert }{\Vert x \Vert } < \infty \, ,
\ee
where the operator norm $\Vert \sfa \Vert $ is determined by the Hilbert space norm $\Vert \sfa x \Vert $; we will denote it with the same symbol by abuse of notation. We will denote by $\cB (\cH)$ the space of all bounded operators acting on $\cH$. For any bounded operator we define its adjoint $\sfa^{\dagger}$ by $\braket{\sfa^\dagger x \vert y} = \braket{x \vert \sfa y}$. Finally an operator $\mathsf{U}$ is unitary if $\mathsf{U} \mathsf{U}^\dagger = \mathsf{U}^\dagger \mathsf{U} = \mathbf{1}$.

The space $\cB (\cH)$ can be endowed with several topologies, which allow us to say that an operator $\sfa$ converges to another operator $\mathsf{b}$. We will mention here some of the most common for completeness:
\begin{itemize}
\item Norm topology: $\Vert \sfa - \mathsf{b} \Vert \longrightarrow 0$.
\item Strong operator topology: $\Vert \left( \sfa - \mathsf{b} \right) \, x  \Vert \longrightarrow 0$ for every $x \in \cH$.
\item Weak operator topology: $\vert \braket{ x \vert \sfa  y }- \braket{ x \vert \mathsf{b} y }\vert \longrightarrow 0$ for every $x,y \in \cH$.
\end{itemize}
These topologies are oriented from the strongest to the weakest, so that an operator convergence in the norm topology implies convergence in the strong operator topology and in the weak operator topology. 

An operator $\sfa \in \cB (\cH)$ is self-adjoint if $\sfa^\dagger = \sfa$ and a projection if $\sfa^2 = \sfa = \sfa^\dagger$. Furthermore an operator $\sfa$ is positive, denoted by $\sfa \ge 0$, if $\braket{x \vert \sfa x} \ge 0 $ for every $x \in \cH$.

An antilinear operator $\sfa$ is defined by $\sfa (x + y) = \sfa x + \sfa y$ and $\sfa (\lambda x) = \overline{\lambda} \sfa x$ for $\lambda \in \complex$. An important example of an antilinear operator is the operator of complex conjugation $J$. Indeed every antilinear operator is of the form $J \sfa$ for some linear operator $\sfa$.

In the study of operator algebras one often encounters unbounded operators. In this case a useful notion is the one of a closed operator. If we denote with $\cD (\sfa) \subset \cH$ the domain of the operator $\sfa$, then we say that the operator $\sfa$ is closed if for every sequence $\{ x_k \}$ such that both $x_k \longrightarrow x$ and $\sfa x_k \longrightarrow v$ we have that $v \in \cD (\sfa)$ and that $\sfa  x = v$ (so that $\sfa x_k \longrightarrow \sfa  x$). An equivalent characterization is via the graph of the operator. The latter is defined as
\be
\Gamma = \left\{ \left( x \, , \sfa \, x \right) \ : \ x \in \cD (\sfa) \right\} \subset{\cH \times \cH} \, .
\ee 
Then one can show that an operator is closed if and only if its graph is a closed subspace of $\cH \times \cH$. We say that an operator is closable if it can be extended to a closed operator on a larger domain.

If we have a self-adjoint operator $\sfa$ we can define the one-parameter group of unitary operators given by $\mathsf{U} (t)= \e^{\ii \sfa \, t}$. The converse also holds, given a (strongly continuous) one-parameter group of unitary operators $\mathsf{U} (t)$, then $\mathsf{U} (t) = \e^{\ii \sfa \, t}$ with $\sfa$ self-adjoint. 

A fundamental result in the theory of operator algebras is the spectral theorem. To begin with consider the case of a self-adjoint operator with a discrete spectrum. Let $\sfa$ be a self-adjoint operator. We define its spectrum as the set of all $\lambda \in \real$ so that the operator $\sfa - \lambda \mathbf{1}$ is not an invertible operator in $\cB (\cH)$, or it or its inverse fail to be bounded. If the operator is also positive, then its spectrum lies in $[0 , \infty )$. The spectral theorem states that the operator can be written as $\sfa = \sum_n \lambda_n \, P_n$ where $\{ \lambda_n \}$ is the discrete spectrum and $P_n$ a family of projections. 

In the more general case, denote by $\sigma (\sfa)$ the spectrum of $\sfa$. Then the spectral theorem states that for any self-adjoint operator $\sfa \in \cB (\cH)$ there exist spectral projections $\mathsf{P} (\lambda)$ so that
\be
\sfa = \int_{\sigma (\sfa)} \lambda \, \dd \mathsf{P}(\lambda) \, .
\ee
A more intuitive way of stating the theorem is as
\be
\braket{\eta \vert \sfa \, \xi} = \int_{\sigma (\sfa)} \lambda \, \dd \!  \braket{\eta \vert  \mathsf{P}(\lambda)  \, \xi } \, .
\ee
The main consequence of this theorem is that given a reasonable (technically Borel-measurable) function $f$ on $\sigma (\sfa)$, we have that
\be
f (\sfa) = \int_{\sigma (\sfa)} f ( \lambda ) \, \dd \mathsf{P}(\lambda)
\ee
is also in $\cB (\cH)$. The theorem allows us to define and use in computations functions of operators. The interested reader can find a more detailed discussion of the spectral theorem in the Appendix \ref{spectral}.

\subsection{States, operator algebras and representations}

The algebra of observables plays a prominent role in quantum physics. Here we will only review the aspects which are relevant to us. In modern language such algebras capture information theoretic aspects of quantum systems and their subsystems, in a sort of model independent way. They can always be thought of as algebras of operators acting on some Hilbert space.

Consider an algebra $\cA$ of bounded operators acting on a Hilbert space $\cH$. We will always assume that our algebras have an identity element. The algebra $\cA$ is called a $C^*$-algebra if $\cA = \cA^{\dagger}$ ( it is closed under taking the adjoint) and it is closed in the operator norm topology. For example a concrete model for an abelian $C^*$-algebra is the algebra of continuous functions over a locally compact space.

The algebra $\cA$ is called a von Neumann algebra if $\cA = \cA^{\dagger}$ ( it is self adjoint) and it is closed in the weak operator topology. This implies that a von Neumann algebra is a $C^*$-algebra, albeit the converse is not true.

An alternative characterization of von Neumann algebras is as follows. Consider the algebra of bounded operators acting on a Hilbert space $\cB (\cH)$. Consider a set $\cM \subset \cB (\cH)$. We define its commutant
\be
\cM' = \left\{ \sfa \in \cB (\cH) \ : \ \left[ \sfa , \mathsf{b} \right] = 0 \ \forall \, \mathsf{b} \in \cM \right\} \ .
\ee
The bicommutant theorem states that a self-adjoint subalgebra $\cA$ of $\cB (\cH)$ is a von Neumann algebra if it is equal to its bicommutant, $\cA = \cA^{''} = \left( \cA' \right)'$.

In the case $\cM$ is a subset of $\cB (\cH)$ consisting of self-adjoint operators, then $\cM'$ is a von Neumann algebra and $\cM''$ is the smallest von Neumann algebra containing the set $\cM$. 

A \textit{factor} is a von Neumann algebra with $\cA \cap \cA' = z \mathbf{1}$ with $z \in \complex$. In words a factor is a von Neumann algebra whose center consists of scalar multiples of the identity operator. Factors are the building block for the classification of von Neumann algebras. For example in finite dimensions factors are always isomorphic to the algebra of $n \times n$ matrices $M_{n} (\complex)$.

A von Neumann algebra $\cA$ is \textit{hyperfinite} if it is generated (as a von Neumann algebra) by an increasing sequence of finite dimensional subalgebras, for example matrix algebras $M_n (\complex)$. This means that $\cA = \left( \cup_n M_n (\complex) \right)''$. Hyperfinite algebras can be approximated by matrix algebras and are the ones of interest in quantum physics.

On a $C^*$-algebra we can define states. A state $\omega$ on a $C^*$-algebra $\cA$ is a continuous linear functional on $\cA$ which is positive and normalized to one, $\omega (1) = 1$. The set of states is convex, so that if $\omega_1$ and $\omega_2$ are states on an algebra, also $\lambda \omega_1 + (1-\lambda) \omega_2$ is a state for all $\lambda \in (0,1)$. The extremal elements of this set, those which cannot be expressed as weighted sums of other states, are called pure states.

A state on a $C^*$-algebra $\cA$ is called \textit{faithful} if $\omega (\sfa^\dagger \sfa) = 0$ if and only if $\sfa = 0$. It is called normal if there is a density matrix, a positive trace-class operator on $\cH$ with $\Tr \rho = 1$ such that
\be
\omega (\sfa) = \Tr_\cH \rho \, \sfa \, .
\ee
Here $\Tr_\cH$ is the trace on the Hilbert space and a trace-class operator is an operator for which this trace is finite. 

A representation of a $C^*$-algebra is a pair $(\cH , \pi)$ of a Hilbert space and a morphism $\pi \, : \, \cA \, \longrightarrow \cB (\cH)$ which preserves the $C^*$-algebra structure (that is preserves the algebra structure and $\pi (\sfa)^\dagger = \pi (\sfa^\dagger)$). The representation is called faithful if $\pi (\sfa) = 0$ implies $\sfa = 0$, and it is called irreducible if it cannot be decomposed into the direct sum of representations. Furthermore a representation is called cyclic, and denoted by the triple $(\cH , \pi , \Omega)$, if there exists a vector $\Omega \in \cH$ such that $\Vert \Omega \Vert = 1$, $(\cH , \pi)$ is a representation and $\pi (\cA) \Omega$ is dense in $\cH$. In this case the vector $\Omega$ is called a cyclic vector.

Any state $\omega$ on the algebra induces a canonical representation, the Gelfand-Naimark-Segal (GNS) representation, which is unique up to unitary equivalence. The GNS representation $(\cH_\omega , \pi_\omega , \Omega_\omega)$ is a cyclic representation such that 
\be \label{GNSdef}
\omega (\sfa) = \braket{\Omega_\omega \vert \pi_\omega (\sfa) \vert \Omega_\omega} 
\ee
for every $\sfa \in \cA$. The converse is also true: any cyclic representation $(\cH , \pi , \Omega)$ defines a state $\omega$ on the algebra $\cA$ by \eqref{GNSdef}.

\subsection{von Neumann algebras, traces and projections} 

The classification of factors is one of the main results of the theory. Such a classification is obtained by studying the traces which one can define on the algebras. A trace is a positive linear functional $\Tr : \cA \longrightarrow \complex$ such that
\be
\Tr (\sfa \sfb) = \Tr (\sfb \sfa) \, .
\ee
For example in finite dimension $\Tr \, \sfa = \sum_i \braket{i \vert \sfa \vert i}$ is the standard trace. In general a trace has the following properties. It is faithful: given a positive operator $\sfa \in \cA^+$ then $\Tr \sfa^\dagger \sfa  = 0$ implies $\sfa = 0$. The trace is semi-finite: for every nonzero $\sfa \in \cA^+$ there is a nonzero $\sfb$ with $\sfb \le \sfa$ and a finite trace. Finally the trace is normal: $\Tr (\sup \sfa_n) = \sup \Tr (\sfa_n)$ for any sequence $\{ \sfa_n \}$. One can show that a trace that is faithful, semi-finite and normal is unique up to rescaling. Therefore one can classify factors by classifying the possible values of traces on the algebra.

A way to do so is to study the possible values of the trace on projections. Recall that a projection is an operator for which $\sfp^2 = \sfp$ and $\sfp^\dagger = \sfp$. If $\sfp$ and $\sfq$ are projections in a von Neumann algebra $\cA$, we say that $\sfp \preceq \sfq$ if there is a partial isometry $v \in \cA$ such that $\sfp = v v^\dagger$ and $v^\dagger v \le \sfq$\footnote{The notation $\sfp \le \sfq$ ($\sfp < \sfq$) means that the range of $\sfp$ is (strictly) contained in the range of $\sfq$. Equivalently we can say that $\braket{x \vert \sfp  x} \le \braket{x \vert \sfq x }$ for every vector $x \in \cH$}. The relation $\preceq$ is a partial order on (the equivalence classes of) projections. In particular if the partial isometry is such that  $\sfp = v v^\dagger$ and $u^\dagger u = \sfq$, we say that $\sfp \approx \sfq$, which is an equivalence relation. We say that a projection $\sfp$ is infinite if $\sfp \approx \sfq$ where $\sfq \lneq \sfp$, and finite otherwise. A von Neumann factor is called infinite if the identity is infinite, and finite otherwise. Finally a projection $\sfp \neq 0$ is minimal if for every projection $\sfq \in \cA$, $\sfq \le \sfp$ implies that $\sfq = \sfp$ or $\sfq = 0$.

With these definitions in place, we can state the classification of von Neumann factors as follows. Consider a factor $\cA$. Then
\begin{itemize}
\item $\cA$ is of type $\mathrm{I}$ if there is a minimal projection. Type $\mathrm{I}$ factors are of the form $\cB (\cH)$ for some $\cH$ and are therefore classified by the dimension of the Hilbert space. If $\dim \cH = n$, with $n \in  \{ 1 , 2 , \dots , \infty \}$, we have a type $\mathrm{I}_n$ factor. These algebras are the algebras of observables which appear in finite and infinite dimensional nonrelativistic quantum mechanics.
\item $\cA$ is of type $\mathrm{II}$ if there is a finite projection but no minimal projections. In particular we say that it is of type $\mathrm{II}_1$ if the identity is finite, and $\mathrm{II}_\infty$ otherwise. In the type $\mathrm{II}_1$ case the trace of projections can assume every value in $[0,1]$ and in the case of $\mathrm{II}_\infty$ factors it can take any value in $[0 , \infty]$. A type $\mathrm{II}_\infty$ factor is always of the form $\cM \otimes \cB (\cH)$ where $\cM$ is a $\mathrm{II}_1$ factor and $\dim \cH = \infty$. Type $\mathrm{II}_1$ factors are not classified. These algebras play a role in quantum gravity and are the main subject of this review. 
\item $\cA$ is of type $\mathrm{III}$ if there is a no finite projection. In particular the trace of projections is infinity (or zero). In practice this means that one cannot define a trace and in particular one cannot define density matrices. These algebras arise in every quantum field theory when studying local operators.
\end{itemize}

All these algebras have a qubit construction, obtained by multiplying an infinite number of appropriate low dimensional quantum systems. See \cite{Witten:2018zxz} for detailed examples.

\subsection{Quantum dynamical systems and KMS states} 

Operator algebras are particularly useful when studying the thermodynamic limit of quantum systems. Abstractly one defines a quantum dynamical system as a pair $(\cA , \alpha)$ where $\cA$ is a von Neumann algebra and $\real \ni t \rightarrow \alpha^t$ a one-parameter group of $*$-automorphisms of $\cA$. This group represents the dynamics and determines time evolution. It is defined via the formal series
\be
\alpha^t \left( \sfa \right) = \sum_{m=0}^\infty \frac{t^m}{m!} \delta^m \sfa = \e^{t \delta} \sfa
\ee
where $\delta$ is the infinitesimal generator of $\alpha$ and $\sfa \in \cA$. The generator enjoys the following two properties: $\delta (\sfa \, \sfb) = \delta (\sfa) \, \sfb + \sfa \, \delta (\sfb)$ (derivation) and $\delta (\sfa^\dagger) = \delta (\sfa)^\dagger$.

We can find a concrete example in the case of a finite dimensional quantum systems with Hamiltonian $H$, where time evolution is given by
\be
\alpha^t (\sfa) = \e^{\ii t H} \, \sfa \, \e^{- \ii t H} \, .
\ee
In this case the infinitesimal generator is $\delta (\sfa) = \ii \left[ H , \sfa \right]$.

In many applications one has a simple dynamics (for example, free dynamics) which can be studied exactly and one is interested in adding a perturbation. The system now evolves according to the perturbed dynamics generated by
\be
\delta_V = \delta + \ii \left[ V , \ \cdot \ \right]
\ee
where $V \in \cA$ is the perturbation operator. If we set $\alpha^t_V = \e^{t \delta_V}$ then we can control the perturbed evolution via the Dyson expansion
\be
\alpha^t_V (\sfa) = \alpha^t (\sfa) + \sum_{n=1}^\infty \int_0^{t} \dd t_1 \int_0^{t_1} \dd t_2 \cdots \int_0^{t_{n-1}} \dd t_{n} \ii \left[ \alpha^{t_n} (V) , \ii \left[ \cdots , \ii \left[ \alpha^{t_1} (V) , \alpha^t (\sfa) \right] \cdots \right] \right] \, .
\ee

An important class of states in quantum dynamical systems is thermal equilibrium states. These are characterized by the KMS condition, named after Kubo-Martin-Schwinger. Before stating this condition, we consider a finite dimensional system. The Gibbs state is defined by
\be
\omega (\sfa) = \frac{1}{Z} \Tr \left( \e^{-\beta H} \sfa \right)
\ee
with $Z = \Tr \left( \e^{-\beta H}\right)$ and we assume $\beta > 0$. Introduce the correlation function
\be \label{KMSfunction}
\cF_\beta (\sfa , \sfb ; t) = \omega (\sfa \, \alpha^t (\sfb)) \, .
\ee
By using the properties of the trace we find
\be
\omega (\sfa \, \alpha^{t} (\sfb)) = \frac{1}{Z} \Tr \left( \e^{- \ii (t - \ii \beta) H} \sfa \e^{\ii t H} \sfb \right) \, .
\ee
Now by analytically continuing $t \longrightarrow t + \ii \beta$
\be
\frac{1}{Z} \Tr \left( \e^{- \ii t H} \sfa \e^{\ii (t + \ii \beta) H} \sfb \right) = \omega (\alpha^t (\sfb) \, \sfa) \, .
\ee
We conclude that the function \eqref{KMSfunction} for $\beta > 0$ is analytic within the strip defined by
\be
S_\beta = \left\{ z \in \complex \, \vert \, 0 < \mathrm{Im} (z ) < \beta \right\},
\ee
where these correlators are convergent if $H$ is only bounded from below; and furthermore it takes the values
\begin{itemize}
\item $\cF_\beta (\sfa , \sfb ; t) = \omega (\sfa \, \alpha^t (\sfb))$,
\item $\cF_\beta (\sfa , \sfb ; t + \ii \beta) = \omega (\alpha^t (\sfb) \, \sfa)$,
\end{itemize}
on its boundary. This is the KMS condition and characterizes thermal equilibrium states even if they are not of the Gibbs form or even if the density matrix does not exist.

\subsection{Modular theory and entropies}

Modular theory is a deep formalism which allows us to study von Neumann algebras without ever making reference to density matrices. Consider a von Neumann algebra $\cA$. Assume $\ket{\Psi}$ is a vector in the Hilbert space on which the algebra is acting. We assume it is cyclic (which means that $\sfa \ket{\Psi}$ is dense and therefore we can generate the whole Hilbert space by acting on it) and separating (which means that $\sfa \ket{\Psi} = 0$ implies $\sfa = 0$). A vector which is both cyclic and separating is referred to as modular in the literature. 

It is convenient to have in mind the finite dimensional case to unpack these definitions. In that case a cyclic and separating vector can be described by a density matrix which has full rank for the algebra and its commutant. Physically the vector has enough entanglement to be able to represent the whole algebra.

We define the Tomita operator
\be
S_{\Psi} \sfa \ket{\Psi} = \sfa^\dagger \ket{\Psi} \, ,
\ee
which is antilinear ($S_\Psi c \ket{\Phi} = \overline{c} S_\Psi \ket{\Phi}$) and unbounded. This operator admits the polar decomposition
\be
S_\Psi = J_\Psi \Delta_\Psi^{1/2}
\ee
in an antiunitary $J_\Psi$ and a $\Delta_\Psi$ positive part. $J_\Psi$ is called the modular conjugation. In particular  $S_\Psi^\dagger S_\Psi = \Delta_\Psi$ plays the role of the modulus of the operator. 

In the case of finite dimensional factors, the positive part can be written in terms of the density matrix of the state $\Psi$ as $\Delta_\Psi = \rho_\Psi (\rho_\Psi')^{-1}$, where $\rho'$ is in the commutant algebra. In general this is not true but we can still define the modular operator. Since this operator is positive we can take its logarithm. We set $\Delta_\Psi = \e^{-h_\Psi}$ where $h_\Psi$ is called the modular Hamiltonian.

The fundamental result is that for $\sfa \in \cA$, $\sfa_s = \e^{\ii s h_\Psi} \sfa \e^{- \ii s h_\Psi}$ remains an element of the algebra $\cA$. The modular conjugation $J_\Psi \sfa J_\Psi$ sends it to an element of the commutant $\cA'$. In particular $\Delta_{\Psi} \ket{\Psi} = 0$ and $J_{\Psi} \ket{\Psi} = \ket{\Psi}$.

Another fundamental property of modular theory is that correlation functions are thermal with respect to the modular hamiltonian. We can see this via the KMS condition
\be
\braket{\Psi \vert \alpha^s (\sfa) \, \sfb \vert \Psi} = \braket{\Psi \vert \sfb \, \alpha^{s + \ii} (\sfa) \vert \Psi} \, .
\ee
Equivalently we can write
\be
\braket{\Psi \vert \sfa \, \sfb \vert \Psi} = \braket{\Psi \vert \sfb \, \Delta_\Psi \, \sfa \vert \Psi} \, ,
\ee
which one can check by writing $\alpha^{s + \ii} (\sfa) $ in term of the modular operator and using that $h_\Psi$ annihilates the state $\Psi$.

All of the above definitions can be generalized to define relative modular operators. The relative Tomita operator is defined as
\be
S_{\Phi \vert \Psi} \sfa \ket{\Psi} = \sfa^\dagger \ket{\Phi} \, .
\ee
Also this operator has a polar decomposition $S_{\Phi \vert \Psi}  = J_{\Phi \vert \Psi}  \Delta_{\Phi \vert \Psi}^{1/2} $. 

In the example of finite-dimensional systems, we have $\Delta_{\Phi \vert \Psi} = \rho_\Phi (\rho_\Psi')^{-1}$. As before one can take the logarithm since the operator is positive and set $\log \Delta_{\Phi \vert \Psi} = - \log \rho_\Phi + \log \rho_\Psi'$. The relative modular operator has the fundamental property that
\be
\braket{\Phi \vert \sfa \, \sfb \vert \Phi} = \braket{\Psi \vert \sfb \, \Delta_{\Phi \vert \Psi} \sfa \vert \Psi} \, ,
\ee
which follows from its definition $\Delta_{\Phi \vert \Psi} = S^\dagger_{\Phi \vert \Psi} S_{\Phi \vert \Psi}$. From the relative modular operator one can define the relative entropy
\be
S_{rel} (\Phi \Vert \Psi) = \braket{\Phi \vert h_{\Psi \vert \Phi} \vert \Phi} = -  \braket{\Phi \vert \log \Delta_{\Psi \vert \Phi} \vert \Phi} \, .
\ee
The relative entropy can be understood as a measure of how much the two states can be distinguished. One can see that $S_{rel} (\Phi \Vert \Psi) \ge 0$ and it is zero iff $\Phi$ and $\Psi$ describe the same state. It follows directly from the definition that the relative entropy is not symmetric. The relative entropy is also monotonic under algebra inclusions: it decreases as we restrict to subalgebras, because we have fewer operators to detect how the operators are different. We stress that since no reference is made to traces of density matrices the relative entropy is well defined also for type $\mathrm{III}$ algebras, as is the case of local operators in  quantum field theory.

If the type of algebra allows for the definition of density matrices, we can rewrite the relative entropy in form that is more familiar. By using $\braket{\Phi \vert h_{\Phi} \vert \Phi} = 0$ we can write
\begin{align}
S_{rel} (\Phi \Vert \Psi) &= \braket{\Phi \vert h_{\Psi \vert \Phi} - h_\Phi \vert \Phi} \cr
& = \braket{\Phi \vert - \log \rho_{\Psi} + \log \rho'_{\Phi} + \log \rho_\Phi - \log \rho'_\Phi \vert \Phi } \cr
& = \Tr \rho_\Phi \left( \log \rho_\Phi - \log \rho_\Psi \right)
\end{align}

\subsection{Type $\mathrm{III}$ algebras in quantum field theory} 

Let us briefly comment on the structure of local algebras in quantum field theory. As this is not the main topic of this note we refer the reader to \cite{Witten:2018zxz,Haag:1996hvx} for a more in depth discussion. In the case of quantum field theory we define local operators $\phi (x)$ at spacetime points $x$. It turns out that these are not really operators but operator-valued distributions. To obtain an operator we need to smear the field as 
\be
\phi_f = \int \dd^4 x f (x) \phi (x) \, ,
\ee
where the test function $f (x)$ is typically chosen from the space of smooth functions with compact support and is supported on some region $\cU$. Now we take bounded functions of this operator (such as $\e^{\ii s \, \phi_f}$) since bounded operators naturally form an algebra (they can be multiplied without worrying about their domain). Finally by taking the weak closure we define the von Neumann algebra $\cA (\cU)$. The latter procedure can be neatly justified since if we have a collection of operators $\sfa_n$ whose matrix elements converge to the matrix element of some operator $\sfa$, then when $n$ is large enough, no experiment can distinguish between $\sfa_n$ and $\sfa$, as discussed in \cite{Haag:1996hvx}.

In the algebraic approach the full information about the theory is contained in the vacuum correlation functions
\be
W^{(n)} (x_1 , \dots x_n) = \braket{\Omega \vert \phi (x_1) \dots \phi (x_n) \vert \Omega} \, .
\ee
Similarly we can define correlations of smeared operators
\be
\braket{\Omega \vert \phi_{f_1} \cdots \phi_{f_n} \vert \Omega} = \int \dd x_1 \cdots \dd x_n f (x_1) \cdots f (x_n) W^{(n)} (x_1 , \cdots , x_n) \, .
\ee
We are glossing over several details here, but in general one has to impose certain analytical conditions \cite{Witten:2018zxz,Haag:1996hvx}.

An important condition is that operators supported in smaller regions give rise to smaller algebras, in the sense that $\cU_1 \subset \cU_2$ implies $\cA (\cU_1) \subset \cA (\cU_2)$. Furthermore causality implies that operators supported in spatially separated regions should commute (and an analog statement for fermions): if $\cU'$ is the causal complement of $\cU$ then $\cV \subset \cU'$ implies that $\cA (\cV) \subset \cA (\cU)'$. Another important result is Haag duality. For a region $\cU$ we form the causal complement $\cU'$. Then the full causal diamond including $\cU$ is $\cU''$, the causal completion of $\cU$. Then we have that $\cA (\cU) = \cA (\cU'')$. Haag duality states that $\cA (\cU') = \cA (\cU)'$, meaning the commutant algebra of $\cU$ is the algebra of its causal complement $\cU'$. This relation is believed to hold in many physical cases, see \cite{Haag:1996hvx,Witten:2018zxz}. An implication of this duality is that the vacuum state in quantum field theory is both cyclic and separating.

Another important result in the theory is the Reeh-Schlieder theorem, which states that the vectors $\phi_{f_1} \cdots \phi_{f_n} \ket{\Omega}$ are dense in the Hilbert space. This implies that the vector $\Omega$ is cyclic for $\cA (\cU)$. In words acting with local operators in a local region of spacetime we can approximate an arbitrary state, even if its support is outside of $\cU$. However this construction is not implemented by a unitary operator.

As an important application, let us consider Rindler space. Consider $W = \{ x^\mu \vert - (x^0)^2 + x^2 \ge 0 \}$, the so-called Rindler wedge. According to the Bisognano-Wichmann theorem, the Minkowski vacuum $\Omega$ restricted to the Rindler wedge $W$ appears thermal with respect to Lorentz boosts. The modular Hamiltonian for this region is the generator of boosts leading to the thermal behavior, which is closely related to the Unruh effect.

For the Rindler wedge the modular operator is $\e^{- 2 \pi K}$ where $K$ is the boost generator. Here $K$ has a continuum spectrum, equal to all of $\real$, since it is a non-compact generator in the Lorentz group. It is a non-trivial fact that a continuous spectrum for the modular operator of the vacuum state is a property that characterizes type $\mathrm{III}$ algebras. Note that we expect every physical state to resemble the vacuum in the UV. This means that the leading short distance contribution to any correlator in any quantum state is given by the operator product expansion and is independent on the particular state we are considering. More specifically the so-called hyperfinite $\mathrm{III}_1$ factor is believed to universally describe the local operator algebras in all quantum field theories.

\subsection{Type $\mathrm{II}_1$ factors and their subfactors}

We have seen that a type $\mathrm{II}_1$ factor is characterized by the fact that every projection is finite but there is no minimal projection. It has a unique trace, up to rescaling. To get a handle on type $\mathrm{II}_1$ factors, we will now discuss an example.

Let $\Gamma$ be a discrete group. Recall that its group algebra is defined as
\be
\complex \Gamma = \left\{ 
\sum_{g \in \Gamma}  \alpha_g \delta_g \ \bigg\vert \  \alpha_g \in \complex \ \text{and $\alpha_g \neq 0$ for finitely many $g$} \right\} \, ,
\ee
where $\delta_g$ is another notation for $g$ which will be more convenient in defining left and right representations. In infinite dimension the group algebra can be completed to a Hilbert space
\be
l^2 (\Gamma) = \left\{ 
\sum_{g \in \Gamma}  \alpha_g \delta_g \ \bigg\vert \ \sum_{g \in \Gamma} | \alpha_g |^2 < \infty \right\} \, ,
\ee
where the inner product $\braket{\delta_g , \delta_h}$ is $1$ if $g = h$ and zero otherwise.

By setting $\lambda (g) \delta_h = \delta_{gh}$ we can define the left regular representation $\lambda \ : \ \complex \Gamma \longrightarrow \cB (l^2 (\Gamma))$ as
\be
\sum \alpha_g \delta_g \longrightarrow \sum \alpha_g \lambda (g)
\ee
on finite sums. Similarly for $\rho (g) \delta_h = \delta_{h g^{-1}}$ we define the right regular representation by
\be
\sum \alpha_g \delta_g \longrightarrow \sum \alpha_g \rho (g)
\ee
again on finite sums. We can now define the group von Neumann algebras $L (\Gamma)$ and $R (\Gamma)$ as the completions of $\lambda \left( \complex \Gamma \right)$ and $\rho \left( \complex \Gamma \right)$ respectively, in the strong operator topology. One can see that they are the commutant of each other, $L \left( \Gamma \right)' = R \left( \Gamma \right)$ and $R \left( \Gamma \right)' = L \left( \Gamma \right)$.

Moreover $L \left( \Gamma \right)$ and $R \left( \Gamma \right)$ are factors iff  for every $h \in \Gamma$ not equal to the identity, each conjugacy class $\left\{ g h g^{-1} \, \vert \, g \in \Gamma \right\}$ of $\Gamma$ is infinite (a condition which ensures that the centre is trivial). 

A notable example is when $\Gamma = S_{\infty} = \bigcup_{n \in \mathbb{N}} S_n$ with $S_n$ the permutation group of $n$ elements. In this case $\mathcal{R} = L (S_{\infty})$ is called the hyperfinite $\mathrm{II}_1$ factor. This is the unique hyperfinite type $\mathrm{II}_1$ factor up to isomorphisms, in the sense that every hyperfinite $\mathrm{II}_1$ factor is isomorphic to $\cR$. 

To any type $\mathrm{II}_1$ factor $\cA$ with trace $\tau \, : \, \cA \longrightarrow \complex$ we can associate the standard representation, which is the GNS representation where the Hilbert space is $L^2 (\cA)$ (the completion of $\cA$ with respect to the inner product $\braket{x , y} = \tau (y^\dagger x)$). In general we can have more complicated representations. We will call a representation of $\cA$ an $\cA$-module. We can form different representations, larger or smaller than the standard representation. For example we can pick a projection $\sfp \in \cA'$ and take $\cH = \sfp L^2 (\cA)$ and obtain a smaller module, or to produce a larger module we can take the tensor product $\cH = l^2 (\mathbb{N}) \otimes L^2 (\cA)$.

In these constructions it is important to realize that the size of $\cA$ and $\cA'$ depend on the particular module they are acting upon. In particular we are interested in understanding if there is a vector $\Omega$ which is both cyclic and separating. Heuristically having a cyclic vector tells us that $\cA$ is rather large, while a separating vector tells us that $\cA'$ is rather large. Interesting representations are those where both $\cA$ and $\cA'$ are big enough to provide a vector which is both cyclic and separating. 

A way to compare the relative size of $\cA$ and $\cA'$ is the \textit{coupling constant} introduced by Murray and von Neumann as follows. One takes an arbitrary vector $\eta \in \cH$ and considers the projections $\sfp$ onto the completion of $\cA \eta$, and $\sfq$ onto the completion of $\cA' \eta$. Then the coupling constant, or $\cA$-dimension of $\cH$ is
\be
\dim_\cA \cH = \frac{\tr_\cA \sfq}{\tr_{\cA'} \sfp} \, .
\ee
In particular one can see that $\dim_\cA \cH = 1$ iff $\cA$ has a cyclic and separating vector.

Consider now $\cB \subset \cA$ both $\mathrm{II}_1$ factors. We define the Jones index of $\cB$ in $\cA$ as
\be
\left[ \cA : \cB \right] = \dim_\cB L^2 (\cA) \, .
\ee
In general $\left[ \cA : \cB \right] \ge 1$ with $\left[ \cA : \cB \right]  = 1$ iff $\cA = \cB$.

If $\left[ \cA : \cB \right] < 4$ then $\cB' \cap \cA = \complex \, \mathbf{1}$ and in this case we call the subfactor $\cB$ \textit{irreducible}. A striking result by Jones states that the possible values for the index are
\begin{itemize}
\item $\left[ \cA : \cB \right] \ge 4$ \, ,
\item $\left[ \cA : \cB \right] = 4 \cos^2 \left( \frac{\pi}{n} \right)$ for $n=3,4,5,\dots$
\end{itemize}
Therefore the index assumes a series of discrete values accumulating up to 4 and after that assumes continuous values. These results are of fundamental importance in the theory of von Neumann algebras and were instrumental in the definition of topological invariants of knots \cite{Jones:1983kv,Jones:1985dw}.

A way to characterize a subfactor is via the conditional expectation. As before given $\cB \subset \cA$, both unital, we define the map $E \, : \, \cA \longrightarrow \cB$ to be a projection onto $\cB$
\be
E ( \sfx ) = \sfx \, , \qquad \forall \, \sfx \in \cB
\ee
which is also $\cB$-linear
\be
E \left( \sfx \, \sfa \, \sfy \right) = \sfx \, E (\sfa) \, \sfy \, , \qquad \forall \, \sfx, \sfy \in \cB \ \text{and} \ \forall \, \sfa \, \in \cA \, .
\ee
In particular the conditional expectation is a completely positive and trace preserving map. A famous result by Umegaki \cite{ume}, states that there exists a unique conditional expectation compatible with a faithful normal trace $\tau$, that is $\tau \circ E = \tau$.

For every $\sfx \in \cA$ we have
\be
E (\sfx) \, e_\cB = e_\cB \, \sfx \, e_\cB
\ee
where $e_\cB \, : \, L^2 (\cA) \longrightarrow L^2 (\cB)$ is the orthogonal projection in $\cB (L^2 (\cA))$. In words the orthogonal projection $e_\cB$ completely determines the conditional expectation.

The subfactor $\cB$ can be characterized by its basic construction. Let us assume that $\left[ \cA : \cB \right] < \infty$. To begin with we define 
\be
\cA_1 = \left\{ \cA \cup \left\{ e_\cB \right\} \right\}'' \subset \cB \left( L^2 (\cA) \right) \, .
\ee
The algebra $\cA_1$ is called the basic construction for $\cB$. It is usually denoted by $\braket{\cA , e_\cB}$ and is the algebra generated by $\cA$ and the projection $e_\cB$.

We can repeat the basic construction and find
\be
\cB \subset \cA \subset \cA_1 \subset \cA_2 \, ,
\ee
where now $\cA_1 = \braket{\cA , e_\cB}$ and $\cA_2 = \braket{\cA_1 , e_\cA}$. One can check that
\begin{align}
e_\cA \, e_\cB \, e_\cA &= \lambda e_\cA \, , \cr
e_\cB \, e_\cA \, e_\cB &= \lambda e_\cB \, ,
\end{align}
with $\lambda = \left[ \cA : \cB \right]^{-1}$.

The iteration of this construction defines the Jones' tower of subfactors:
\be
\cB \subset \cA  \stackrel{e_{1}}{ \subset}\cA_1  \stackrel{e_{2}}{ \subset}\cA_2  \stackrel{e_{3}}{ \subset}\cA_3 \cdots 
\ee 
In the tower each factor is defined by induction as $\cA_{i+1} = \braket{\cA_i , e_{i+1}}$, where we have set $e_{i+1} \equiv e_{\cA_{i+1}} \, :  L^2 \left( \cA_{i+1} \right) \longrightarrow  L^2 \left(\cA_i \right)$. 

Remarkably these projections satisfy the relations of the Temperley-Lieb algebra
\begin{align}
e_i^2 &= e_i = e_i^\dagger \cr
e_i \, e_j & = e_j \, e_i \ , \qquad \text{if} \ \ | i- j | > 2 \cr
e_i \, e_{i \pm 1} \, e_i &= \lambda \, e_i
\end{align}
where again  $\lambda = \left[ \cA : \cB \right]^{-1}$. In particular for any word $w$ in the letters $\{ e_1 , \cdots , e_n \}$ we have that $\tau (w e_{n+1}) = \lambda \tau (w)$.

This is the origin of the famous relation between type $\mathrm{II}_1$ factors and knots. A knot $\hat{b}$ can be realized as the closure of a braid $b$. The braid group $\mathsf{B}_n$ is the group generated by the elements $\{ \sigma_1 , \cdots , \sigma_{n-1} \}$ with relations
\begin{align}
\sigma_i \, \sigma_j & = \sigma_j \, \sigma_i \ , \qquad \text{if} \ \ | i- j | > 2 \cr
\sigma_{i+1} \, \sigma_i \, \sigma_{i+1} & = \sigma_i \, \sigma_{i+1} \, \sigma_i \, .
\end{align}
If we denote the Temperley-Lieb algebra generated by $\{ e_1 , \cdots , e_n \}$ by $\mathsf{TL}_n (\lambda)$, then we can define the representation $\rho_t \, : \, \mathsf{B}_n \longrightarrow \mathsf{TL}_n (\lambda)$ by
\begin{align}
\rho_t (1) & = 1 \cr
\rho_t (\sigma_i) & = 1 - (1 + t) e_i \cr
\rho_t (\sigma_i^{-1}) & = 1 - (1 + \frac{1}{t})  e_i 
\end{align}
with $\lambda^{-1} = 2 + t + \frac{1}{t}$.

Consider now a link $\hat{b}$ obtained from the closure of a braid $b$. We can apply the map $\rho_t$ to $b$ to obtain an element of the Temperley-Lieb algebra. The Jones polynomial of $\hat{b}$ is proportional to the trace $\tau \left( \rho_t (b) \right)$ taken in the type $\mathrm{II}_1$ factor \cite{Jones:1985dw}.

\subsection{The crossed product} \label{CPintro}

The crossed product is a key construction in the theory of operator algebras, which in particular turns a type $\mathrm{III}_1$ algebra into a type $\mathrm{II}$ algebra \cite{takesaki}. This construction was first applied to quantum gravity in \cite{Witten:2021unn}. A modern introduction to the topic can be found in \cite{Sorce:2024pte} or in the appendix of \cite{Kudler-Flam:2023qfl}.
 
Consider a type $\mathrm{III}_1$ algebra $\cA$ with $\Psi$ a cyclic and separating vector. Let $\Delta_\Psi$ be the associated modular operator and $h_\Psi$ the modular Hamiltonian. To define the crossed product one introduces an auxiliary Hilbert space $L^2 (\real)$ and the associated algebra of bounded operators. Consider two operators, $p$ and $x$ acting on $L^2 (\real)$ which we can think of as momentum and position.

The crossed product is then defined as the von Neumann algebra generated by
\be \label{cp1}
\cA \rtimes \real = \braket{ \sfa \otimes 1 , \e^{\ii h_\Psi s} \otimes \e^{\ii p \, s}  \, \vert \, \sfa \in \cA , s \in \real} \, .
\ee
One of the main results of \cite{takesaki} is that if $\cA$ is of type $\mathrm{III}_1$ then $\cA \rtimes \real$ is of type $\mathrm{II}_\infty$. In this case the automorphism generated by the modular Hamiltonian becomes inner.

An equivalent expression for the crossed product can be obtained by using the commutation theorem \cite{daele}, which gives
\be \label{cp2}
\cA \rtimes \real = \{ \widehat{\sfa} \in \cA \otimes \cB (L^2 (\real)) \, \vert \, \left[ h_\Psi - x  , \widehat{\sfa} \right] = 0 \, .
\}
\ee
When expressed in this fashion the crossed product selects elements of the extended algebra $\cA \otimes \cB (L^2 (\real))$ which commute with the constraint $h_\Psi - x $.

Note that the forms \eqref{cp1} or \eqref{cp2} of the crossed product appears to depend explicitly on the vector $\Psi$ which is used to construct the modular Hamiltonian. However, this is not the case, and different vectors give rise to isomorphic algebras (see for example \cite{Witten:2021unn}).

\section{Gravitational algebras}  \label{GA} 

In this Section we will introduce gravitational algebras, von Neumann algebras of type $\mathrm{II}$ which enter in the study of perturbative quantum gravity in certain backgrounds.

\subsection{A type $\mathrm{III}$ algebra}

To begin with we consider the eternal black hole in AdS. The authors of \cite{Leutheusser:2021qhd,Leutheusser:2021frk} used the boundary theory to define an operator algebra from the large $N$ limit of thermal correlators, above the Hawking-Page temperature. In this limit operators of the form $\cW = \Tr W - \langle \Tr W \rangle_\beta$, that is single-trace operator with their thermal expectation value removed, have non-trivial two-point functions. All the other correlators vanish. Such operators correspond to generalized free fields in the bulk. This operator algebra on the right boundary is a von Neumann algebra $\cA_{0,R}$ of type $\mathrm{III}$. From this algebra one can construct via the GNS construction a Hilbert space which is well defined in the large $N$ limit, starting from the thermofield double state $\Psi$. Its commutant $\cA_{0,L} = \cA_{0,R}'$ is a copy of the same algebra, this time constructed starting from the left boundary. The holographic duality identifies these two algebras with the bulk algebras $\cA_{r,0}$ and $\cA_{l,0}$ \cite{Leutheusser:2021qhd,Leutheusser:2021frk}, governing the dynamics of the quantum fields in the two exteriors of the eternal black hole.  

The two algebras $\cA_{R,0}$ and $\cA_{L,0}$ have trivial centers; they are so-called \textit{factors}. In particular they do not contain the boundary Hamiltonians $H_R$ and $H_L$.  These operators are conserved charges which correspond to the black hole mass by holography. Note that these operators do not have a large $N$ limit but only their difference $\widehat{H} = H_R - H_L$ does. This operator is related to the bulk operator $\widehat{h}$, which generates time translations in the bulk, as $\widehat{h} = \beta H$. To obtain operators which have a large $N$ limit one can define the operators
\be
U_L = \frac{H_L - \braket{H_L}_\beta}{N} \, , \qquad U_R = \frac{H_R - \braket{H_R}_\beta}{N} \, .
\ee 
Note that the two operators $U_L$ and $U_R$ coincide in the strict $N = \infty$ limit; therefore in this limit we can simply call them $U$. This operator is central and we can add it to the algebras $\cA_{0,R}$ and $\cA_{0,L}$ by simply tensoring them with the algebras of bounded functions of $U$.

\subsection{The crossed product} \label{CPwitten}

It was shown in \cite{Witten:2021unn} that including $1/N$ corrections amounts to the crossed product construction. Due to the relation $U_R = U_L + \beta \widehat{H} / N$, when we include $1/N$ corrections the operators $U_L$ and $U_R$ become distinct. Therefore $U_R$ is now given by the sum of $U_L$, which commutes with $\cA_{0,R}$, and the modular Hamiltonian, which generates a one-parameter group of automorphisms, the modular flow. Since the algebra is of type $\mathrm{III}$ these automorphisms are outer. We now take $X = \beta N U_L$. The crossed product algebra $\cA_R = \cA_{0,R} \rtimes \real$ now acts on $\widehat{\cH} = \cH \otimes L^2 (\real)$ and is of type $\mathrm{II}_\infty$. In particular since the algebra is of type $\mathrm{II}_\infty$ we can define a trace and therefore density matrices and entropies.

We will consider a particular class of states which take the form $\widehat{\Psi} = \Psi \otimes g (X)^{1/2}$, where $g$ is a Gaussian function. We refer to these states as classical-quantum. For these states it is easy to see that the modular operator has the explicit form \cite{Witten:2021unn}
\be
\widehat{\Delta}_{\widehat{\Psi}} = \Delta_{\Psi} \, g (\beta \widehat{H} + X) \, g (X)^{-1} = K \widetilde{K} \, ,
\ee
where $K \in \cA_R \rtimes \real$ and $\widetilde{K} \in \left( \cA_R \rtimes \real \right)'$ are given by
\begin{align}
K &= \e^{- \left(\beta \widehat{H} + X \right)} g (\beta \widehat{H} + X) \, , \\ 
\widetilde{K} &= \e^X g(X)^{-1} \, .
\end{align}
This factorization can be used explicitly to define a trace on the algebra. For $\widehat{\sfa} \in \cA_R$ we have
\be \label{trK-1}
\tr \, \widehat{\sfa} = \braket{\Psi | \widehat{\sfa} K^{-1} | \Psi} = \int_{- \infty}^{\infty} \dd X \, \e^X \, \braket{\Psi | \sfa (X) | \Psi} \, .
\ee
Note that this trace is only defined up to a rescaling of $K$. Also the trace is not defined on all elements of the algebra; for example it gives infinity on the identity.

Due to the presence of a factor of $N$ in the exponent, inside the operator $X$ this trace is only a formal function of $N$. This does not affect however the computation of the entropies \cite{Chandrasekaran:2022eqq}. We will keep using the expression \eqref{trK-1} as a formal expression.

Now we can use \eqref{trK-1} to define density matrices and entropies. Given a state $\widehat{\Phi}$ we call the operator $\rho_{\widehat{\Phi}} \in \cA_{R}$  a density matrix if we have that 
\be 
\tr \, \widehat{\sfa} \, \rho_{\widehat{\Phi}} = \braket{\widehat{\Phi} | \widehat{\sfa} | \widehat{\Phi}} \, ,
\ee
for every operator $\widehat{\sfa} \in \cA_{R}$. The associated von Neumann entropy in the algebra $\cA_R$ is then
\be
S (\widehat{\Phi})_{\cA_R} = - \braket{\widehat{\Phi} \vert \log \rho_{\widehat{\Phi}} \vert \widehat{\Phi} } \, .
\ee
Note that the entropy defined in such a fashion has an additive ambiguity, much like entropy in classical statistical mechanics.

For example if we consider the classical-quantum state $\widehat{\Psi} = \Psi \otimes g(X)^{1/2}$, it follows from \eqref{trK-1} that $\rho_{\widehat{\Psi}} = K$ and we can write the entropy
\be \label{vnPsi}
S (\widehat{\Psi})_{\cA_R} = \int_{- \infty}^{\infty} \dd X \left( X \, g(X) - g(X) \, \log X \right) \, .
\ee
For more general classical-quantum states $\widehat{\Phi} = \Phi \otimes f (X)^{1/2}$ the entropy can be computed by taking the  expectation value of the formal operator \cite{Chandrasekaran:2022eqq}
\be
h_{\widehat{\Phi}} = - \frac{1}{N} \log \rho_{\widehat{\Phi}} \, .
\ee
The result is that
\be \label{EntropyPhi}
S (\widehat{\Phi})_{\cA_R} = N \beta \braket{U_R} + N S_0 - S  \left( \Phi \Vert \Psi \right) - \braket{\log | f (U_R) |} + \braket{\alpha (U_R)} \, .
\ee
Here one can fix the function $\alpha$ be computing the next correction in the $1/N$ expansion:
\be
\alpha (U_R) = - \frac{N^2}{T^2 \, C_{\rm{BH}}} \frac{U^2_R}{2} + \rm{const} \, .
\ee
The parameter $C_{\rm{BH}}$ is called black hole heat capacity. Note that while the entropy \eqref{EntropyPhi}  depends explicitly on $ N S_0 $ and $\braket{\alpha (U_R)}$, these quantities are state independent and therefore cancel when one computes entropy differences. 

There is an alternative construction based on the microcanonical ensemble where one does not need to use formal arguments  \cite{Chandrasekaran:2022eqq}. However the canonical ensemble is more suited to the study of nonequilibrium dynamics.

\subsection{De Sitter and the hyperfinite $\mathrm{II}_1$ factor}

The second example that we will consider is de Sitter spacetime. Let us consider quantum fields in a fixed de Sitter background. Such a spacetime has a natural vacuum state for the quantum fields, the Bunch-Davies state $\Psi_{\mathrm{dS}}$, defined via analytical continuation from the Euclidean theory. Without dynamical gravity, correlation functions of quantum fields have a thermal interpretation in an ensemble with inverse temperature $\beta_{\mathrm{dS}}$. This picture is altered when including perturbative dynamical gravity.

In a spacetime with closed spatial slices it is problematic to impose the gravitational constraint. A possible way out is to add an external observer which can be used to define operationally the algebra of observables \cite{Chandrasekaran:2022cip}. For a different take based on redefining the Hilbert space inner product see \cite{Higuchi:1991tk,Higuchi:1991tm}. 

Let us consider the static patch. The combined Hamiltonian of the system is now
\be
\widehat{H} = H + q
\ee
where $H$ is the Hamiltonian which generates time translations in the static patch while $q$ is the observer's Hamiltonian\footnote{The reader should not confuse the operators $p$ and $q$ introduced in this Section and associated with the observer, with the general notation for projections $\sfp$ and $\sfq$ used elsewhere in the text.}. We also introduce the conjugate operator $p = - \ii \frac{\dd}{\dd q}$. In this simplest example an observer is just a clock, capable of measuring time. We shall however require that its energy is positive (or more generally bounded from below). In the limit $G_N \longrightarrow 0$ the interactions between the observer and the quantum fields can be neglected. We denote by $\cA_0$ the algebra of quantum fields along the worldline of the observer which acts on a Hilbert space $\cH_0$. For a more detailed discussion of observers and their role in dynamical gravity we refer the reader to \cite{AliAhmad:2024wja,AliAhmad:2024vdw,Chen:2024rpx,DeVuyst:2024pop,Fewster:2024pur,Hoehn:2023ehz,Kudler-Flam:2024psh,Witten:2023qsv,Witten:2023xze}.

The full Hilbert space is the tensor product of $\cH_0$ with the observer's Hilbert space,
\be
\cH = \cH_0 \otimes L^2 (\real_+) \, ,
\ee
and the algebra of observables is obtained by imposing the Hamiltonian constraint
\be
\cA =  \left( \cA_0 \otimes \cB (L^2 (\real_+) \right)^{\widehat{H}} \, .
\ee
As explained in Section \ref{CPintro} this is precisely a crossed product. To understand this better let us forget for the moment about the positivity of the energy of the observer. We claim that the algebra of operators which commute with the constraint $\widehat{H}$ is generated by $\left\{ \e^{\ii p H} \sfa \e^{- \ii p H} , q \right\}$. 

To check this claim we only have to check that operators of the form $\e^{\ii p H} \sfa \e^{-\ii p H}$ commute with $H + q$. Indeed
\begin{align} \label{dScomm-1}
\left[ H + q , \e^{\ii p H} \sfa \e^{-\ii p H} \right] = \e^{\ii p H} \left[ H , \sfa \right] \e^{- \ii p H} + \left[ q , \e^{\ii p H} \sfa \e^{-\ii p H} \right] \, .
\end{align}
To compute the second term we use the fact that since $p = - \ii \frac{\dd}{\dd q}$ generate translations:
\be
\e^{\ii p H} q \e^{- \ii p H} = q + H \, .
\ee
Therefore we must have
\be
\left[ q , \e^{\pm \ii p H} \right] = \mp H \e^{\ii p H} \, .
\ee
Finally since $\left[q , \sfa \right]=0$ since they are element of different algebras
\be
\left[ q , \e^{\ii p H} \sfa \e^{-\ii p H} \right] = \left[ q , \e^{\ii p H} \right] \sfa \e^{-\ii p H} +  \e^{\ii p H} \sfa  \left[ q ,\e^{-\ii p H} \right] = \e^{\ii p H} \left[ \sfa , H \right] \e^{- \ii p H}
\ee
and comparing with \eqref{dScomm-1} we get the desired result. 

One can obtain an equivalent description by conjugating with $\e^{- \ii p H}$. This has the effect of removing the ''dressing'' of the operators $\sfa$ and shifting $q$. Now the algebra is generated by operators of the form $\left\{ \sfa , q - H \right\}$ where now $q - H \ge 0$. A more convenient form is $\left\{ \sfa , H + x \right\}$ with $x = - q$, for which the constraint becomes $H + x \le 0$. We call this algebra $\cA_{cr}$. The physical algebra is obtained from $\cA_{cr}$ by imposing the constraint  $H + x \le 0$. Since $\cA_{cr}$ is obtained from the crossed product of a type $\mathrm{III}$ algebra, it is a type $\mathrm{II}_\infty$ algebra. 

Since $\cA_{cr}$ is constructed from the tensor product $\cA_0 \otimes L^2 (\real)$, an operator $\widehat{\sfa} \in \cA_{cr}$ can be understood as an $\cA$-valued function of $H + x$. When evaluating matrix elements, we will use that $H \ket{\Psi_{\mathrm{dS}}} = 0$. This leads us to define a trace as
\be
\Tr \, \widehat{\sfa} = \int_{- \infty}^{\infty} \beta_{dS} \dd x \e^{\beta_{dS} x} \braket{\Psi_{dS} \vert \sfa (x) \vert \Psi_{dS}} \, .
\ee
Again this trace is not finite on all the elements of the algebra, but when it is well defined, it is positive and cyclic.

Finally to impose the constraint that $q \ge 0$ one introduces the projector $\Pi = \Theta (q)$. With this projection one obtains the physical algebra $\widehat{\cA} = \Pi \cA_{cr} \Pi$ which acts on the Hilbert space $\Pi (\cH \otimes L^2 (\real)) = \cH \otimes L^2 (\real_+)$. The trace is the same trace but now restricted to operators of the form $\Pi \, \widehat{\sfa} \, \Pi$.

It is then easy to check that applying this projection has turned this algebra from a type $\mathrm{II}_\infty$ algebra to a type $\mathrm{II}_1$ algebra, by computing the trace of the identity
\be
\Tr_{\widehat{\cA}} 1 = \int_{- \infty}^{+ \infty} \beta_{\mathrm{dS}} \dd x \e^{\beta_{\mathrm{dS}} x} \braket{\Psi_{\mathrm{dS}} \vert \Theta (- H - x) \vert \Psi_{\mathrm{dS}}} = 1 \, ,
\ee
where we have used that $H \ket{\Psi_{\mathrm{dS}}} = 0$ so that the effect of the step function is to reduce the integration domain to $[-\infty , 0]$.

In this algebra there is a maximum entropy state $\rho = 1$. This state can be purified as
\be
\Psi_{\mathrm{max}} = \Psi_{\mathrm{dS}} \, \otimes  \sqrt{\beta_{\mathrm{dS}}} \e^{\beta_{\mathrm{dS}} x / 2}
\ee
which physically represents the de Sitter state tensored with a thermal energy distribution (in the sense that $|\Psi_{\mathrm{max}}|^2 \sim \e^{\beta_{\mathrm{dS}} x}$) associated with the observer.

We can compute entropies, at least for a special class of density matrices associated with semiclassical states of the form
\be
\widehat{\Phi} = \Phi \otimes f(x)
\ee
with $f (x) = \sqrt{\epsilon} \, g (x \epsilon)$ and $\epsilon << \beta_{\mathrm{dS}}$. These conditions ensure that this function is slowly varying, allowing the observer to measure events with a time uncertainty much smaller than $\beta_{\mathrm{dS}}$, thus retaining a semiclassical interpretation of spacetime. In this approximation the density matrix associated to a semiclassical state of this form is
\be
\rho_{\widehat{\Phi}} = \frac{1}{\beta} \overline{f} (x + h_\Psi / \beta) \e^{- \beta x} \Delta_{\Phi \vert \Psi} f (x + h_\Psi / \beta) + \cO (\epsilon) \, ,
\ee
and the associated entropy is given by
\begin{align}
S (\rho_{\widehat{\Phi}}) &= - \Tr \, \rho_{\widehat{\Phi}} \log \rho_{\widehat{\Phi}} = - \braket{\widehat{\Phi} \vert \log \rho_{\widehat{\Phi}} \vert \widehat{\Phi}} \cr
&= - \braket{\Phi \vert h_{\Psi \vert \Phi} \vert \Phi} + \braket{\widehat{\Phi} \vert h_\Psi + \beta x \vert \widehat{\Phi}} + \int_{-\infty}^0 \dd x | f (x) |^2 \left(
- \log |f (x)|^2 + \log \beta \right) \, .
\end{align}
Here the first term is the relative entropy between the state $\Phi$ and the de Sitter state. The second term is the expectation value of the energy of the observer. Finally the last term represents the entropy in the observer's energy fluctuations. Putting all these terms together one obtains the generalized entropy \cite{Chandrasekaran:2022cip}
\be
S_{\mathrm{gen}} = \frac{A}{4 G_N} + S_{\mathrm{out}} \, .
\ee

\section{Nonequilibrium dynamics of finite quantum dynamical systems}  \label{NE}

In this Section we will review some aspects of the nonequilibrium dynamics of finite quantum dynamical systems, i.e. quantum systems governed by a type $\mathrm{I}$ algebra. In our exposition we have found the reviews \cite{JP0,JP1,RuelleRev,strasberg} particularly useful. We will see in the next Section how these statements generalize to gravitational algebras.

\subsection{Generalities}

We will start with some general ideas and comments. Classical thermodynamics is characterized by the phenomenological observation that certain states of matter, the equilibrium states, can be completely described by a handful of functions, called state functions. Equilibrium statistical mechanics, one of the triumphs of modern physics, can under certain circumstances reproduce these functions and the laws governing their behaviour from an analysis of the microscopic dynamics. 

Nonequilibrium physics is however not understood, both in classical and quantum mechanics. Perhaps the main reason is that while there is a certain universality governing equilibrium physics, there are several physically distinct behaviors pertinent to out-of-equilibrium dynamics. Because of this nonequilibrium physics has been traditionally concern with system close to equilibrium, where powerful fluctuation-dissipation theorems are available.

However many interesting phenomena take place far from equilibrium. The last few decades have seen remarkable progress in our understanding of physics far from equilibrium, for example with the introduction of general fluctuation theorems.

A general setup to study general features of nonequilibrium physics is to consider a small system and couple it to reservoirs. A reservoir is a large (or infinite) system in equilibrium at fixed temperature, typically consisting of free particles. The interaction with the system is localized at an interface, so that the degrees of freedom within the reservoir which are influenced by the interaction, move to infinity in the reservoir and can be forgotten. In other words the thermodynamical behaviour of the reservoirs is not influenced by the original system.

The simplest nonequilibrium states engineered in this fashion are called nonequilibrium steady states, and we will discuss them momentarily. They are stationary states which still describe a non trivial transfer of energy or particles. A typical observable for nonequilibrium steady states is the rate of entropy production. While general thermodynamics quantities are defined only at equilibrium, entropy production makes sense also far from equilibrium and is generally used to study nonequilibrium steady states.

Even far from equilibrium one can obtain exact results. In the past decades a series of fluctuation relations holding far from equilibrium were discovered, starting with the Jarzynski equality \cite{jarzynski}. These relations go collectively by the name of fluctuation theorems and form by now a vast and active research field; see \cite{strasberg} for a review. Essential to fluctuation theorems is the time-reversal invariance of the microscopic dynamics. Fluctuation theorems then are generally relations which relate the probability of a process with the probability of the time-reversed process. For example by considering an isolated system in equilibrium at inverse temperature $\beta$, one can compare the probability that a certain work $W$ is performed on the system by an external time-dependent driving force with the probability that a work $-W$ is performed by the time-reversed external force. The Jarzynski equality 
\be
\braket{\e^{- \beta W}} = \e^{- \beta \Delta F}
\ee
relates this work with the difference $\Delta F$ of free energy between the initial equilibrium state and the final equilibrium state.

\subsection{Nonequilibrium steady states}

Equilibrium states in thermodynamics can be operationally defined by specifying certain state functions, such as temperature and entropy. There is no explicit reference to the dynamics, regardless of the fact that the dynamics is needed to specify the microscopic ensembles.

On the other hand the dynamics is essential to understand out-of-equilibrium systems. To this date we are very far from a comprehensive understanding of physics out of equilibrium. The simplest situation is provided by nonequilibrium steady states (NESS), which are those steady states where the system settles after imposing a forcing, given by an external field or a steep gradient of thermodynamic parameters. For example one can imagine putting into contact two systems at different temperatures, creating a temperature gradient. As a result one will create fluxes of the extensive quantities used to parametrize the system, such as the energy. These fluxes will determine a non-trivial rate of entropy production.

To be concrete let us consider a quantum dynamical system $(\cA , \alpha)$ and assume that the system is initially in an $\alpha$-invariant state $\omega$. We use a self-adjoint operator $V \in \cA$ to perturb the system. The resulting perturbed evolution will be denoted by $\alpha_V^t$. Then a NESS is defined via the limit
\be \label{NESSdef}
\omega_+ (\sfa) = \lim_{t_k } \frac{1}{t_k} \int_{0}^{t_k} \, \omega \circ \alpha_V^t (\sfa) \, \dd t,
\ee
if this exist, with $\{ t_k \}_{k \in \zed_+}$ a divergent sequence. In words this definition describes a stationary state where the system settles after the perturbation. If this is not an equilibrium state -- which would be the case if the perturbation is sufficiently small -- then it describes a genuine nonequilibrium state.

If the perturbation is small enough the system will settle into a new stationary state $\omega_V$ which is KMS with respect to the perturbed evolution $\alpha_V$. This is a standard statement, discussed for example in \cite{BR1,BR2}. To avoid this situation one needs a different setup.

A common way to engineer NESS is to couple the system to external reservoirs. We introduce a collection of reservoirs $\{ \cR_1 , \cdots ,\cR_M \}$ collectively denoted by $\cR$, modelled after quantum dynamical systems $(\cO_{\cR_i} , \alpha_{\cR_i})$. Every reservoir has its own algebra of observables $\cO_{\cR_i}$, its own evolution operator $\alpha_{\cR_i}$ and it is assumed to be in thermal equilibrium at  inverse  temperature $\beta_i$. These equilibrium states are described by the $\alpha_{\cR_i}$-invariant KMS states $\omega_i$.

We couple the system to the reservoirs by a perturbation $V = \sum_{j=1}^M V_j$ where $V_j = V_j^\dagger \in \cA \otimes \cO_{\cR_{j}}$ describes the interaction between the system and the reservoir $\cR_j$. If the perturbation is chosen appropriately the system will settle in a nonequilibrium state. We can expect this state to be characterized by non trivial fluxes, describing the exchange of energy between the system and the reservoir.

Note that the reservoirs are assumed to be infinte systems at fixed temperature. In particular this means that they have an internal chaotic dynamics. The latter acts as a source of randomness for the original system which will also become chaotic.

\subsection{Entropy production}

Nonequilibrium dynamics is usually associated with entropy production \cite{strasberg}. In the same setup as above, where the system is coupled to external reservoirs at inverse temperature $\beta_i$, we expect a steady flow of heat through the system. In  any stationary state, the entropy flux entering or leaving the system will determine the rate of entropy production.

Consider now a finite dimensional system and let us denote by $H_S$ its Hamiltonian. If we consider a stationary state, the energy leaving the reservoirs, represented by the operator
\be
- \sum_k \beta_k \Phi_k \, ,
\ee
determines the rate of entropy flux into the system.

Let us describe the interaction between the system and the reservoir by the Hamiltonian
\be \label{Hsysres}
H = H_S + V + \sum_k H_{\cR_k} \, .
\ee
Then the Heisenberg equation \cite{JP1} determines the energy flux as
\be \label{kflux}
\Phi_k = - \ii \left[ H , H_{\cR_k} \right]  = \delta_k (V) \, .
\ee
Furthermore we denote with $\delta_k = \ii \left[ H_{\cR_k}, \, \cdot \, \right]$ the generator of the dynamics of the reservoirs. We assume that the reservoirs are sufficiently big that their thermal equilibrium is not altered by the interaction.

We can rewrite the above expression in the language of quantum dynamical systems as follows. Consider a state $\mu$. We define the entropy production in this state as
\be \label{EPdef}
\mathrm{Ep} \left( \mu \right) = \mu \left(
- \sum_{k=1}^M \beta_k \, \delta_k (V) 
\right) \, .
\ee
This expression allows for a straightforward extension to systems with infinite dimension.

To better illustrate the situation, let us consider the case of a finite dimensional system, which we imagine divided into subsystems, as in  \cite{RuelleRev}. Then the observables in this systems are elements of a type $\mathrm{I}$ von Neumann algebra consisting of bounded operators acting on a finite dimensional Hilbert space. Let $\rho (t)$ be the density matrix of the whole system. This operator may depend on time; however since the overall system is isolated, the von Neumann entropy remains constant over time. We can introduce partial density matrices $\rho_a$ by tracing over degrees of freedom outside the $a^{\rm th}$ subsystem. The corresponding von Neumann entropies $S (\rho_a (t))$ can now vary with time. In this setup  \cite{RuelleRev} defines the rate of entropy production as
\be \label{EPru}
e = \frac{\dd}{\dd t} \left( \sum_a S (\rho_a (t) ) - S \left( \rho (t) \right) \right) \, .
\ee
In particular the subadditivity property of entropy guarantees that the expression in parentheses is positive. This expression represents the information that we lose about $\rho$ when we partition the system into subsystems. The quantity in parentheses represents the rate of change in the entropy of each subsystem. One can verify by direct computation that \eqref{EPru} agrees with \eqref{EPdef}.

\section{Quantum thermodynamics of gravitational algebras} \label{QT}

In this Section we will outline the generalization of the nonequilibrium formalism described in Section \ref{NE} to the theory of gravitational algebras. We refer the reader to \cite{Cirafici:2024jdw,Cirafici:2024ccs} for the technical details.

\subsection{Nonequilibrium dynamics and entropy production}

We now return to the eternal black hole in AdS as discussed in Section \ref{BHAdS}. We want to couple the left and right algebras to a collection of reservoirs and then take the crossed product to include gravitational corrections.

After introducing an interaction term the total Hamiltonians are given by
\begin{align} \label{Htot}
H_{\mathrm{tot},R} = H_R + H_{\omega,R} + V_R \, ,  \\ \nonumber
H_{\mathrm{tot},L} = H_L + H_{\omega,L} + V_L \, .
\end{align}
In this expression we have introduced the two self-adjoint operators  $V_R$ and $V_L$ which represent the interaction between the original systems and the reservoirs. Explicitly
\be
V_R = \sum_{j=1}^M V_{R,j} \, ,
\ee
where $V_{R,j} \in \cA_{R} \otimes \cO_{b_j,R}$, and similarly for $V_L$. We require that the reservoirs on the left and right boundaries are conjugate to each other. Furthermore we require that the algebra $\cA_{0,R} \otimes_V \cO_{b,R}$ is a Type $\mathrm{III}_1$ algebra and we denote by $H_\omega$ the Hamiltonians of the reservoirs.

To begin with let us assume that the perturbed state is KMS. Such a state can be constructed by first considering the decoupled theory and then adding the interaction. It is also necessary that all the reservoirs are at the same temperature. A short computation shows that this state has the form
\be
\Psi_V = \e^{-\beta (\widehat{H} + \widehat{H}_\omega + V)/2} \left( \Psi \otimes \Omega_\omega \right) \, .
\ee
In particular this implies that the vector $\Psi_V$ is both cyclic and separating. We can construct the associated modular operator as
\be
\Delta_{\Psi_V} = \e^{- \beta L_V} \, ,
\ee
where
\be
L_V =  \widehat{H} + \widehat{H}_\omega + \widehat{V} \, ,
\ee
and $\widehat{V} = V_R - V_L$.

Since we know the modular operator, we can determine the dynamics $\tau_V$ of the system
\be
\tau_V^s \left( \sfa_V \right) = \e^{\ii s \, L_V} \sfa_V \e^{- \ii s \, L_V}  \, ,
\ee
for any $\sfa_V \in \cA_{0,R} \otimes_V \cO_{b,R}$. Here we employ the notation $\otimes_V$ to stress that now the two algebras are interacting. 

We would like now to incorporate the $1/N$ corrections in this construction. To begin with we set $T = \beta L_V$ and $X = \beta N U_L$. One can see that $\e^{\ii T s}$ generates and outer automorphism for  $\cA_{0,R} \otimes_V \cO_{b,R}$. We can therefore take the crossed product to obtain the algebra $\cA^{(b)}_{R,V} = \left( \cA_{0,R} \otimes_V \cO_{b,R} \right) \rtimes \real$.

Consider the classical-quantum state
\be
\widehat{\Psi}_V = \Psi_V \otimes g(X)^{1/2} \, .
\ee
We can find its modular operator by requiring that
\be
\braket{\widehat{\Psi}_V \vert \widehat{\sfa}_V \widehat{\sfb}_V \vert \widehat{\Psi}_V}
= \braket{\widehat{\Psi}_V \vert \widehat{\sfb}_V \widehat{\Delta}_{\Psi_V} \widehat{\sfa}_V \vert \widehat{\Psi}_V} \, .
\ee
Indeed a short computation gives\be
\widehat{\Delta}_{\Psi_V} =  \Delta_{\Psi_V}  g \left(\beta \, L_V + X \right) \, g (X)^{-1} \, .
\ee

The modular operator has an important property: it factorizes as
\be
\widehat{\Delta}_{\Psi_V} = \widetilde{\cK}_V \, \cK_V \, ,
\ee
where
\begin{align}
\cK_V & = \e^{- \left(\beta L_V + X \right) } g \left( \beta L_V + X \right) = \e^{- \beta \left[ L_V + \frac{X}{\beta} - \frac{1}{\beta} \log g \left( \beta L_V + X \right) \right]} \, , \cr 
\widetilde{\cK}_V & = \frac{\e^X}{g (X)} \, .
\end{align}
Because of this the $*$-automorphism
\be \label{tauVhat}
\widehat{\tau}_V^s \left( \widehat{\sfa}_V \right) = \widehat{\Delta}_{\Psi_V}^{- \ii s/\beta} \, \widehat{\sfa}_V \, \widehat{\Delta}_{\Psi_V}^{\ii s/ \beta} = \cK_V^{- \ii s / \beta} \, \widehat{\sfa}_V \, \cK_V^{\ii s / \beta}
\ee
is now inner. The reason for this is that $\cK_V$ is now an element of algebra $\cA^{(b)}_{R,V}$. This automorphism can be physically interpreted as the modular time evolution of a certain quantum dynamical system that accounts for gravitational corrections in the coupling between gravitational algebras and reservoirs. Explicitly
\be
\widehat{\tau}_V^s \left( \widehat{\sfa}_V \right) = \e^{\ii t \, I_V} \,  \widehat{\sfa}_V \,  \e^{-\ii t \, I_V} 
\ee
where the modular Hamiltonian is given by
\be \label{modH-IV}
I_V = L_V + \frac{X}{\beta} - \frac{1}{\beta} \log g \left( \beta L_V + X \right) \, .
\ee
Now that we have constructed the generator of the dynamics of the system we can use it to study its dynamics out of equilibrium. Note that this interpretation is a consequence of the relationship between the equilibrium statistical weights and the evolution operator, familiar from statistical mechanics.

Finally we can employ this construction to define a trace as in Section \ref{CPwitten}
\be
\tr \,  \widehat{\sfa}_V =\braket{\widehat{\Psi}_V \vert \widehat{\sfa}_V \cK_V^{-1} \vert \widehat{\Psi}_V}   = \braket{\widehat{\Psi}_V \vert \widehat{\sfa}_V \frac{\e^X}{g (X)} \vert \widehat{\Psi}_V} = \int_{-\infty}^{+ \infty} \dd X \, \e^X 
\braket{\Psi_V \vert \widehat{\sfa}_V  \vert \Psi_V} \, .
\ee
This trace is finite for a certain subalgebra, comprising all those operators for which the integral is convergent. Since we know how to define a trace, we can now define density matrices and their von Neumann entropies.

By using the definition of the trace one can also define density matrices and therefore the von Neumann entropy. For example the density matrix of the classical-quantum state $\widehat{\Psi}_V$ is $\cK_V$ itself, since
\be \label{TrPsiV}
\Tr  \, \widehat{\sfa}_V \, \cK_V = \braket{ \widehat{\Psi}_V \vert \cK_V \cK_V^{-1}   \widehat{\sfa}_V  \vert \widehat{\Psi}_V } = \braket{ \widehat{\Psi}_V \vert  \widehat{\sfa}_V  \vert \widehat{\Psi}_V } \, .
\ee
Note that
\be
\cK_V  \log \cK_V = \e^{- \left( \beta L_V + X \right)} g \left( \beta L_V + X \right) \left(
- \left( \beta L_V + X \right) + \log g \left( \beta L_V + X \right)
\right) \, .
\ee
By using the definition of the trace \eqref{TrPsiV} and the fact that $L_V \Psi_V = 0$ we can compute the von Neumann entropy for $\widehat{\Psi}_V$. We find
\be \label{vnPsiV}
S (\widehat{\Psi}_V)_{\cA^{(b)}_{R,V}}  = \int_{-\infty}^{+ \infty} \dd X \left( X g(X) - g(X) \log g(X) \right)
\ee
precisely as in \eqref{vnPsi}! Indeed a weak time independent perturbation which takes an initial thermal state into a new thermal state is not associated with any entropy production and can  be thought of as an adiabatic process. Physically as we perturb the system the ground state change accordingly. However the situation will be very different in the case of time dependent interactions, or if we consider reservoirs with different temperatures.

When the reservoirs have different temperatures, we expect to have a non trivial entropy production. To study entropy production let us pick a reference state with respect to which we will define the production of entropy. Let us now consider $\widehat{\Psi} = \Psi \otimes \Omega_\omega \otimes g(X)^{1/2}$, which is a state of the non-interacting system. The generator of its modular group is 
\be \label{deltapsi}
\delta_{\widehat{\Psi}} = \ii \sum_{k=1}^L \beta_k \left[ \widehat{H}_{\omega_k} , \, \cdot \, \right] + \ii \beta \left[ I , \, \cdot \, \right] \, .
\ee
We have introduced the operator
\be \label{modH-I}
I = \widehat{H} + \frac{X}{\beta} - \frac{1}{\beta} \log g \left( \beta \widehat{H} + X \right) \, .
\ee
Now we couple the boundary theory and the reservoirs and take into account $1/N$ corrections. In finite dimensional systems the entropy production observable is described by the action of the generator of the modular group of the reference state on the interaction term. In this context things are more complicated due to the presence of gravitational corrections and due to the explicit form of the operator $\widehat{\tau}_V$. In order to sidestep these problems, it is natural to consider as interaction the difference between the infinitesimal generators of the coupled and decoupled theories, respectively $\delta_{\widehat{\tau}_V}$ and $\delta_{\widehat{\tau}}$.

We can read the form of $\delta_{\widehat{\tau}_V}$ from \eqref{tauVhat}
\be \label{deltatauVhat}
\delta_{\widehat{\tau}_V} = \ii \left[ I_V \, , \, \cdot \, \right] \, ,
\ee
where $I_V$ is given in \eqref{modH-IV}. To keep things simple let us take a time independent interaction $V$. We introduce the interaction term
\be \label{cVint}
\cV = I_V - I - \sum_{k=1}^M \widehat{H}_{\omega_k} \, ,
\ee
and we define the entropy production observable as
\be \label{defsigmaV}
\sigma_V = - \delta_{\widehat{\Psi}} (\cV) \, .
\ee
This means that the entropy production rate in a state $\widehat{\varphi}$ can be computed as
\be
\mathrm{Ep} (\widehat{\varphi}) =\widehat{\varphi} \left( \sigma_V \right) \, .
\ee
As a particular example consider the vector $\widehat{\Phi}$ which represents a classical-quantum state. Then
\be
\mathrm{Ep} (\widehat{\varphi}) = \widehat{\varphi} (\sigma_V) = \braket{\widehat{\Phi} \, \vert \, \sigma_V \,  \vert \, \widehat{\Phi}} \, .
\ee
Not all states are of this form. In particular we are interested in the case where the state is a NESS
\be
\chi^+ (\sfa) = \lim_{t_k} \frac{1}{t_k} \int_0^{t_k}  \widehat{\psi} \circ \widehat{\tau}^{ \, t}_V (\sfa) \dd t =  \lim_{t_k} \frac{1}{t_k} \int_0^{t_k}  \braket{ \widehat{\Psi}  \, \vert \, \widehat{\tau}_V^{\, t} (\sfa)  \, | \, \widehat{\Psi} } \dd t \, ,
\ee
which is obtained from the thermofield double state $\widehat{\Psi}$. In this case a short computation gives
\be
\mathrm{Ep} (\chi^+) =  - \sum_{j=1}^M \beta_j \, \chi^+ \left( {\mathbf \Theta}_j  \right) \, .
\ee
We have introduced the operators 
\be \label{ThetaFlux}
{\mathbf \Theta}_k =  \delta_{\cR_k} (\cV) = \ii  \left[ \widehat{H}_{\omega_k} , \cV \right] \, ,
\ee
which are formally the analog of \eqref{kflux}. We interpret these operators as the energy flux in/out the reservoir, when also gravitational corrections are present. We see that the formalism incorporates gravitational corrections in accord with the laws of thermodynamics.

Finally it is worth noticing that entropy production is related to the relative entropy. This follows from the relation, proven in  \cite{JP2}, 
\be \label{EPformulaGen}
S \left( \widehat{\Phi}^U  \Vert \widehat{\Psi} \right) =  S \left( \widehat{\Phi}  \Vert \widehat{\Psi} \right)  - \ii \widehat{\Phi} (U^\dagger \, \delta_{\widehat{\Psi}} (U)) \, .
\ee
In this expression we define $\widehat{\Phi}^U $ for a unitary operator $U$, by the property that $\widehat{\Phi}^U (\sfa) = \widehat{\Phi} (U^\dagger \sfa \, U)$ for every  $\sfa \in \cA$.

From \eqref{EPformulaGen} one can obtain the identity
\be
S \left( \widehat{\Phi} \circ \widehat{\tau}^t_V \Vert \widehat{\Psi} \right) =  S \left( \widehat{\Phi}  \Vert \widehat{\Psi} \right)  - \ii \widehat{\Phi} \left( \Gamma^t_V \, \delta_{\widehat{\Psi}} (\Gamma^{t*}_V) \right) \, .
\ee
In this expression $\widehat{\tau}^t$ represents perturbed evolution associated with the interaction. This identity relates entropy production with the relative entropy with respect to a reference state.

Relative entropy is well defined for a purely normal state $\widehat{\Phi}$. However a NESS will not be normal in general. Assume however that we have a NESS $\widehat{\chi}^+$ obtained from  $\widehat{\Psi}$ by a sequence $\{ \widehat{\tau}_V^{t_n} \}_{n \in \zed_+}$. Then one can see that 
\be
\lim_{n \rightarrow \infty} \frac{1}{t_n} S \left(\widehat{\Psi} \circ \widehat{\tau}^{t_n}_V \Vert \widehat{\Psi} \right) = 
\lim_{n \rightarrow \infty} \frac{1}{t_n} \int_0^{t_n} \widehat{\Psi} \circ \tau_V^s (\sigma_V) \, \dd s = \mathrm{Ep} (\widehat{\chi}^+) \, .
\ee
We reach the conclusion that, since relative entropy is non-decreasing,
\be
\mathrm{Ep} (\widehat{\chi}^+) \ge 0
\ee
which expresses the physical fact that in the NESS  $\widehat{\chi}^+$ the entropy production is non-negative.

All the expression discussed so far can be computed more easily in perturbation theory, see \cite{Cirafici:2024jdw} for details.

\subsection{Fluctuation theorems in de Sitter}

In ordinary quantum mechanics an observer can understand static fluctuations of a system by preparing several identical copies of the same system and performing projective measurements of an observable. The result is a probability distribution for the eigenvalues of the observable in a particular state. On the other hand to access \textit{dynamical} fluctuations one must allow for the system to evolve in time after a first measurement before performing a second measurement. This two-time measurement scheme is naturally related to nonequilibrium physics.

We will now discuss how this perspective applies to gravitational algebras. To be concrete we will consider the static patch of de Sitter spacetime along with an observer capable of doing measurements. This setup is accurately described by the hyperfinite type $\mathrm{II}_1$ factor \cite{Chandrasekaran:2022cip}. By extending the two-time measurement scheme to gravitational algebras we will be able to derive fluctuation theorems. This construction is already known in the case of finite-dimensional systems \cite{strasberg}.

To begin with assume that the system is in a semiclassical state $\rho_{\widehat{\Phi}}$. The observer chooses to measure the entropy observable $S = - \log \rho_{\widehat{\Phi}}$. We use the spectral theorem to decompose $S = \sum_{s} s \, \Pi_s$. For simplicity we are assuming that we can use a discrete model, since the observer's instrument has a finite resolution. However all the results can be easily extended to the continuous formalism, for example by using the techniques illustrated in the Appendix \ref{spectral}.

The observer performs a measurement at $t = 0$. If the eigenvalue $s$ is observed, after the measurement the system is in the state
\be
\frac{\Pi_s \rho_{\widehat{\Phi}} \Pi_s}{\Tr  \, \Pi_s \rho_{\widehat{\Phi}}}  \, .
\ee
The observer lets the system evolve for a time $t$ and then performs another measurement. The probability of observing the eigenvalue $s'$ is given by
\be
p (s' , s) = p (s' \vert s) p (s) = 
\Tr \left( \Pi_{s'} 
\e^{- \ii t H} \rho_{\widehat{\Phi}} \Pi_s \e^{\ii t H}
\right) \, .
\ee
A more interesting quantity is the probability of observing an average change of entropy $\overline{s} = \frac{s'-s}{t}$ during the time $t$
\begin{align}
\mathbf{P}_t (\overline{s}) 
& = 
 \sum_{s' , s } \delta \left( \left(s - s'\right) - t \overline{s} \right) 
\braket{\Psi_{\mathrm{max}} \vert
 \Pi_{s'} \ 
\e^{- \ii t H} \rho_{\widehat{\Phi}} \Pi_s
\vert \Psi_{\mathrm{max}}
} \, .
\end{align}
Here we have used the fact that the trace can be expressed via $ \Psi_{\mathrm{max}}$, the maximum entropy state, where $H \Psi_{\mathrm{max}} = 0$, as discussed in \ref{dS}.

Recall that the entropy observable is given by $S = - \log \rho_{\widehat{\Phi}}$. We can use this fact to relate the above probability to the correlator $\Tr  \left( \rho_{\widehat{\Phi}}^{\alpha} \ \tau ( \rho_{\widehat{\Phi}}^{1-\alpha}) \right) $. Indeed by direct computation we see
\begin{align} \label{relRenyi}
\Tr  \left( \rho_{\widehat{\Phi}}^{\alpha} \ \tau ( \rho_{\widehat{\Phi}}^{1-\alpha}) \right) 
&=
\sum_{s , s'} \e^{- \alpha (s' - s)} \, \Tr \left(
\e^{-\ii t H} \rho_{\widehat{\Phi}} \Pi_s \e^{\ii t H} \Pi_{s'}
\right)
\cr & = 
\sum_{s , s'} \e^{- \alpha (s' - s)} \,  
\braket{\Psi_{\mathrm{max}} 
\vert
\e^{-\ii t H} \rho_{\widehat{\Phi}} \Pi_s \e^{\ii t H} \Pi_{s'}
\vert
 \Psi_{\mathrm{max}}}
\, .
\end{align}
Thus we can write
\be
\Tr \left(
 \rho_{\widehat{\Phi}}^{\alpha} \, \tau ( \rho_{\widehat{\Phi}}^{1 - \alpha} ) \right) = \sum_{\overline{s}} \mathbf{P}_t (\overline{s}) \e^{- t \alpha \overline{s}} \, .
\ee
Assuming that the theory and in particular the state $\widehat{\Phi}$ are invariant under time reversal, we find that
\be \label{TRidentity}
\Tr \left(
 \rho_{\widehat{\Phi}}^{\alpha} \, \tau ( \rho_{\widehat{\Phi}}^{1 - \alpha} )
\right) =\Tr \left(
 \rho_{\widehat{\Phi}}^{1-\alpha} \, \tau ( \rho_{\widehat{\Phi}}^{ \alpha} )
\right) \, .
\ee
From this identity one can derive the following fluctuation theorem
\be
\mathbf{P}_t (- \overline{s}) = \e^{- t \overline{s}} \mathbf{P}_t (\overline{s}) \, .
\ee
Note that this expression holds also out of equilibrium, since never in the derivation we assumed thermal equilibrium. Its physical interpretation is that negative entropy fluctuations are exponentially suppressed with respect to positive entropy fluctuations.

We can now try to generalize the above argument to other observables. We will do so by comparing transition amplitudes for a process and the same process time-reversed. Consider an observable $Y$ and its spectral decomposition $Y = \sum_y  y \, \Lambda_y$. The observer implements the two-time measurement protocol: the probability to observe the value $y_0$ at time $t=0$ and the value $y_\tau$ at time $t = \tau$ is
\be
P [y_\tau , y_0] = \Tr \left( \Lambda_{y_\tau} \e^{- \ii \tau H} \Lambda_{y_0} \, \rho_{\widehat{\Phi}} \, \Lambda_{y_0} \e^{\ii \tau H} \Lambda_{y_\tau} \right) \, .
\ee
We wish to compare this probability with that for the time-reversed process. By time-reversed process we intend the process with initial state $ \rho_{\widehat{\Phi}}^{tr} = \e^{- \ii \tau H}  \rho_{\widehat{\Phi}} \e^{\ii \tau H} $ (the time evolved density matrix) and where the evolution is reversed in time, that is $\rho_{\widehat{\Phi}}^{tr}  (t) = \e^{\ii t H}  \rho_{\widehat{\Phi}} \e^{-\ii t H}$.

We associate the transition probability
\be
P^{tr} [y_0 , y_{\tau}] = \Tr \left( \Lambda_{y_0} \e^{ \ii \tau H} \Lambda_{y_\tau} \, \rho_{\widehat{\Phi}}^{\tr} \, \Lambda_{y_\tau} \e^{-\ii \tau H} \Lambda_{y_0} \right)
\ee
to the time reversed process. In order to measure quantitatively how these two probabilities differ, we introduce the quantity
\be
\Xi[y_\tau , y_0] = \log \frac{P [y_\tau , y_0]}{P^{tr} [y_0 , y_{\tau}] } = -  \Xi^{tr} [y_0 , y_\tau] \, .
\ee
which possesses the property
$\braket{\e^{-\Xi}} = \sum_{y_\tau , y_0} P [y_\tau , y_0] \e^{- \Xi [ y_\tau , y_0 ]} = 1
$, 
which by Jensen's inequality $\braket{\e^J} \ge e^{\braket{J}}$, implies $\braket{\Xi} \ge 0$.

To obtain an abstract fluctuation theorem we define the two new quantities 
\begin{align}
p (\Xi) = \sum_{y_{\tau} , y_0} P [y_\tau , y_0] \, \delta \left( \Xi - \Xi[y_\tau , y_0]  \right) \, , \cr
p^{tr} (\Xi) = \sum_{y_{\tau} , y_0} P^{tr} [y_\tau , y_0] \, \delta \left( \Xi - \Xi^{tr} [y_0 , y_\tau]  \right) \, ,
\end{align}
which measure the probability of attaining a certain value of $\Xi$ for the forward and backward process. The fluctuation theorem then reads
\begin{align} \label{abstractFT}
p (\Xi) &=  \sum_{y_{\tau} , y_0} P^{tr} [y_\tau , y_0]  \, \e^{\Xi [y_{\tau} , y_0 ]} \, \delta \left( \Xi - \Xi[y_\tau , y_0]  \right)  \cr
 & = \e^{\Xi} \sum_{y_{\tau} , y_0} P^{tr} [y_\tau , y_0]  \, \delta \left( \Xi  +  \Xi^{tr}[y_0 , y_\tau]  \right) \cr
 & = \e^{\Xi} p^{tr} (-\Xi) \, .
\end{align}
To see an application of this result, let us assume that our state $\rho_{\widehat{\Phi}}$ admits the coarse-grained spectral decomposition
\be \label{cgrho}
\rho_{\widehat{\Phi}} = \sum_y \frac{p_y}{d_y} \Lambda_y \, ,
\ee
where $p_y = \Tr \rho_{\widehat{\Phi}} \Lambda_y$ and $d_y$ is given by $\Tr \Lambda_y = d_y$. The parameter $d_y \in [0,1]$ is characteristic of type $\mathrm{II}_1$ algebras and corresponds to the continuous dimension of the projection. In our formalism it is necessary to ensure the correct normalization of the density matrix. Physically we interpret it as corresponding to the number of states associated with the observer value $y$. The fact that it is not an integer is a consequence of the renormalization of the trace, which in  type $\mathrm{II}_1$ algebras is finite because an infinite constant has been subtracted.

By using the ansatz \eqref{cgrho} we can compute directly
\begin{align} \label{logPs}
\log \frac{P [y_\tau , y_0]}{P^{tr} [y_0 , y_{\tau}] } & = \log \frac{\Tr  \rho_{\widehat{\Phi}} \Lambda_{y_0}}{\Tr  \rho_{\widehat{\Phi}} \Lambda_{y_\tau}}  + \log \frac{d_{y_\tau}}{d_{y_0}} \cr
& = \left[ S (\rho_{y_0}) - S (\rho_{y_\tau}) \right] - \left(\Tr \rho_{y_0} \mathscr{H} - \Tr \rho_{y_\tau} \mathscr{H}  \right)  + \log \frac{d_{y_\tau}}{d_{y_0}} \, .
\end{align}
In this expression we have introduced $\mathscr{H} = - \log \rho_{\widehat{\Phi}}$ and defined the normalized density matrix
\be
\rho_y = \frac{ \e^{-\mathscr{H}} \Lambda_y }{\Tr  \e^{-\mathscr{H}} \Lambda_y} \, .
\ee
The terms in \eqref{logPs} have distinct physical interpretations compared to those of standard quantum thermodynamics. The first term determines which process is thermodynamically favored, since it compute an entropy difference. The second term captures the difference in the modular Hamiltonian's expectation values in the two projected states.

Similar to quantum thermodynamics, the nonequilibrium dynamics can be described in terms of equilibrium quantities. The second term can counterbalance the entropy change from the first term, potentially increasing the likelihood of a process that would otherwise be disfavored. The last term is unique to the structure of $\mathrm{II}_1$ algebras: it is a consequence of the prescription to implement gravitational constraints by involving an observer. Note that since the dimension of the projections $d_{y}$ can approach zero, this term may dominate the first two. Therefore we find the prediction that some processes, though entropically suppressed, could still be favored due to this offset. 

\section{Conclusions}

In this review, we have provided a concise overview of the theory of gravitational algebras and its applications to nonequilibrium physics. The main takeaway is that Lorentzian perturbative quantum gravity knows that the Bekenstein-Hawking entropy, or more accurately the generalized entropy, has a statistical interpretation. This arises because the structure of the algebra of observables is fundamentally altered by the inclusion of gravitational effects, even at the perturbative level. When the algebra of observables is of type $\mathrm{II}$, we can use the properties of the algebra to define density matrix operators and to compute their von Neumann entropies. While absolute entropies are still not physical, entropy differences are.

Aspects of gravitational nonequilibrium physics can be studied from the perspective of the quantum statistical mechanics of type $\mathrm{II}$ algebras, a largely unexplored subject. We have focused on two such aspects: the coupling of the theory to external reservoirs to induce nonequilibrium steady states with non-trivial entropy production, and fluctuation theorems. For other topics not covered here, we refer the reader to \cite{Cirafici:2024jdw,Cirafici:2024ccs} for further details. For example, studying quantum channels in de Sitter space is closely related to the theory of subfactors of the hyperfinite type $\mathrm{II}_1$ factor, with a similar connection existing in black hole physics \cite{vanderHeijden:2024tdk,AliAhmad:2024saq}. A different perspective on nonequilibrium aspects of gravitational algebras was discussed in \cite{Kudler-Flam:2023qfl,Chen:2024rpx,Kudler-Flam:2024psh} and it would be interesting to understand better the relation between the two points of view.

One of the main messages of this review is that the theory of gravitational algebras should be viewed as akin to quantum stochastic thermodynamics -- a quantum statistical theory that, while not fully microscopic, is suited to describing mesoscopic systems. This aligns with the fact that type $\mathrm{II}$ algebras have no irreducible representations, and therefore no microstates. Nevertheless, the theory appears to capture the essential thermodynamic properties of spacetimes with horizons and may offer important insights into the mysteries of quantum gravity.

\vskip0.5cm
 \noindent {\bf {Acknowledgements.}} 
 I thank the INFN for financial support via the Iniziativa Specifica GAST. I am a member of the IGAP and of the INDAM-GNFM.
\begin{appendix} 
 
\section{The Spectral Theorem} \label{spectral}

In this Appendix we will discuss a few more details about the spectral theorem. We refer the reader to \cite{BR1} for more details.

Let us begin by considering the finite dimensional case. Let $\cH$ be a finite dimensional Hilbert space and we consider a self adjoint operator $\sfa = \sfa^\dagger$ on $\cH$. The operator $\sfa$ can be represented, upon choosing a basis, by a Hermitian matrix, which for simplicity will be denoted by the same symbol. Such a matrix can be diagonalized by a unitary transformation, its eigenvalues are real and eigenvectors corresponding to distinct eigenvalues are orthogonal. Recall from elementary algebra that the eigenvalues $\lambda$ are solutions of the characteristic equation
\be \label{chareq}
\det (\sfa - \lambda \mathbf{1}) = 0
\ee
and the corresponding eigenvectors obey $\sfa \, u = \lambda u$. The Hilbert space decomposes as $\cH = \oplus_i \cH_i$ where
\be \label{Hker}
\cH_i = \ker  (\sfa - \lambda \mathbf{1}) 
\ee
which for simplicity we can assume to be one-dimensional (simply counting the eigenvalues with their multiplicity). Then every eigenspace $\cH_i$ defines a projection $\sfp_i$. Then the spectral theorem states that
\begin{align} \label{st-fin}
\sfa = \sum_i \lambda_i \, \sfp_i \cr
\mathbf{1} = \sum_i \sfp_i \, .
\end{align}
An immediate consequence of \eqref{st-fin} is that we can define functions of the operator $\sfa$ as
\be \label{fc-fin}
f ( \sfa ) = \sum_i f ( \lambda_i ) \, \sfp_i  \, .
\ee
The set of the eigenvalues of $\sfa$ is called the spectrum of the operator. In the finite dimensional case it is a finite discrete set. From \eqref{chareq} it can also be defined as the set where the operator $\sfa - \lambda \mathbf{1}$ fails to be invertible. This way of thinking extends more easily to infinite dimensions.
 
 Let us now consider the case where the Hilbert space is infinite dimensional and we have a bounded operator $\sfa \in \cB (\cH)$. We say that $\lambda \in \complex$ is an element of the resolvent set $\rho (\sfa)$ if the operator $\sfa - \lambda \mathbf{1}$ is invertible in $\cB (\cH)$. The latter condition requires that $\sfa - \lambda \mathbf{1}$ is injective in $\cH$, with inverse densely defined in $\cH$ and bounded. The complement $\sigma (\sfa)$ of $\rho (\sfa)$ in $\complex$ is called the spectrum of $\sfa$. It usually comprises a discrete part $\sigma_d (\sfa)$,  when $\sfa - \lambda \mathbf{1}$ is not injective, a continuous part $\sigma_c (\sfa)$, where $\sfa - \lambda \mathbf{1}$ is densely defined over $\cH$ but is not bounded, and a residual part $\sigma_r (\sfa)$, where $\sfa - \lambda \mathbf{1}$ is not densely defined. The latter part is absent in many cases of interest (for example for self-adjoint operators) and we will ignore it. A value $\lambda \in \sigma_d (\sfa)$ is an eigenvalue of the operator $\sfa$ to which we associate an eigenvector. On the other hand $\lambda \in \sigma_c (\sfa)$ is not an eigenvalue since no eigenvector exists. To circumvent this problem it is customary in quantum mechanics to enlarge the space of functions to include distributions; in this case one can talk about generalized eigenvectors.
 
 Many operators of interest are not bounded or defined over the whole Hilbert space. This is the case for the position and momentum operators in quantum mechanics. In this case the theory is more difficult and one has to impose certain conditions (for example require that the operators are densely defined and closed) to ensure the existence of the adjoint operator.
 
 We will now focus on bounded operators $\sfa \in \cB (\cH)$. Within bounded operators there is a very simple class, the compact operators, which can be characterized\footnote{This is actually a theorem, the original definition of a compact operator is that the closure of the image of every bounded set is a compact set.} as bounded operators with finite rank (that is $\dim (\mathrm{Im} \, \sfa)) < \infty$ ). 
 
 The spectral theorem for self-adjoint compact operators is very similar to the case of finite dimensions, since there is no continuous part. For a compact self-adjoint operator $\sfa$ there exists a complete orthonormal basis made of eigenvectors. This fact can be used to construct the decomposition \eqref{Hker} and the spectral theorem has the form \eqref{st-fin}. Functional calculus can be also set up as in \eqref{fc-fin}.

For more general self-adjoint operators there will also be a continuous spectrum. In this case the spectral theorem has the form
\be
\sfa =  \sum_{\sigma_d (\sfa)} \lambda_i \, \sfp_i + \int_{\sigma_c (\sfa)} \lambda \,  \dd \mathsf{P} (\lambda) \, .
\ee
 It is notationally convenient to group together the discrete and the continuous part and write simply
 \be
 \sfa = \int_{\sigma (\sfa)} \lambda \,  \dd \mathsf{P} (\lambda) \, .
 \ee
 To explain the meaning of these expression we have to define the measure $\dd \mathsf{P}$. This is a measure defined on $\sigma (\sfa)$ which takes values in $\cB (\cH)$, that is it associates to every subset $\Delta \in \sigma (\sfa)$ a projection $\mathsf{P} (\Delta)$.
 
 To explain how to define this measure let us recall a few basic results on measure theory. A $\sigma$-algebra on a set $X$ is a non-empty collection $\Sigma$ of subsets of $X$, which contains $X$ itself and is closed under complement, countable unions and countable intersections. A positive measure is a function $\mu \, : \, \Sigma \longrightarrow [0 , \infty]$ which is additive on countable unions $\mu (\bigcup_\alpha u_\alpha) = \sum_\alpha \mu (u_\alpha)$, where the $u_\alpha$ are disjoint sets. A complex-valued measure can be constructed from four positive measures.
 
 A spectral measure (or projection valued measure) is a map $\mathsf{P} \, : \, \sigma \longrightarrow \cB (\cH)$ so that $\mathsf{P} (\emptyset) = 0$, $\mathsf{P} (X) = \mathbf{1}$, $\mathsf{P} (u)$ is an orthogonal projection and the following two relations hold
\begin{align}
\mathsf{P} (u \cap u') &= \mathsf{P} (u) \mathsf{P} (u') \cr
\mathsf{P} (u \cup u') &= \mathsf{P} (u) + \mathsf{P} (u') \, , \qquad \text{if} \ u \cap u' = \emptyset
\end{align}
For every $x,y \in \cH$
\be
\mathsf{P}_{xy} (u) = \braket{\mathsf{P} (u) \, x \, \vert \, y}
\ee
 is a complex-valued measure.
 
 Then the spectral theorem states that every self-adjoint $\sfa \in \cB (\cH)$ determines uniquely a spectral measure $\mathsf{P}$ on $\sigma (\sfa)$ so that
  \be
 \sfa = \int_{\sigma (\sfa)} \lambda \,  \dd \mathsf{P} (\lambda) \, .
 \ee
 This expression should be taken to mean
  \be
 \braket{ \sfa x \vert y } = \int_{\sigma (\sfa)} \lambda \,  \dd \mathsf{P}_{xy} (\lambda) \,
 \ee
 for every $x,y \in \cH$. Furthermore
\be
 \mathbf{1} = \int_{\sigma (\sfa)}  \dd \mathsf{P} (\lambda) \, .
\ee
Finally the functional calculus can be defined by
\be
f ( \sfa ) = \int_{\sigma (\sfa)} f (\lambda ) \,  \dd \mathsf{P} (\lambda) \, .
\ee
for every bounded function $f$. Equivalently
\be
\braket{ f (\sfa ) x \vert y } = \int_{\sigma (\sfa)} f (\lambda ) \,  \dd \mathsf{P}_{xy} (\lambda) \, .
\ee

 \end{appendix}

\end{document}